\documentclass[aps,10pt,prd,twocolumn,superscriptaddress,preprintnumbers,eqsecnum,%
showkeys,nofootinbib,showpacs]{revtex4-2}
\usepackage{bm}
\usepackage{graphicx}
\usepackage[T1]{fontenc} 
\usepackage{slashed}
\usepackage{mathrsfs}  
\usepackage{amsmath}
\usepackage{amssymb}
\usepackage{enumerate}
\allowdisplaybreaks
\usepackage{xr-hyper}
\usepackage{hyperref}
\hypersetup{colorlinks=true,citecolor=blue,linkcolor=blue,urlcolor=blue}
\newcommand{\secondpart}{Part II~\cite{CW_kin_deut}}
\usepackage{orcidlink}
\begin{document}
\title{Semi-inclusive deep-inelastic scattering on a polarized spin-1 target. \\
I.\ Cross section and spin observables}
\author{W.~Cosyn\,\orcidlink{0000-0002-9312-8569}}
\email[ E-mail: ]{wcosyn@fiu.edu}
\affiliation{Department of Physics, Florida International University, Miami, Florida 33199, USA}
\author{C.~Weiss\,\orcidlink{0000-0003-0296-5802}}
\email[ E-mail: ]{weiss@jlab.org}
\affiliation{Theory Center, Jefferson Lab, Newport News, VA 23606, USA}
\begin{abstract}
We develop the theoretical framework for semi-inclusive deep-inelastic scattering on a polarized spin-1 target
and apply it to scattering on the polarized deuteron with spectator nucleon tagging.

In Part I (this article) we present the general form of the semi-inclusive cross section
and polarization observables for the spin-1 target. A relativistically covariant formulation in terms
of 4-vectors and invariant polarization parameters is employed. The target polarization is described
by a spin density matrix with vector and tensor polarization. The spin and azimuthal angle dependence
of the semi-inclusive cross section is derived and parametrized in terms of invariant structure functions.
To validate the result, the structure functions are expressed as photon-target helicity amplitudes
with known symmetry properties. The expressions presented here are kinematic (no assumptions about
particle production dynamics) and valid in all regions of the deep-inelastic final state (current
and target fragmentation regions).

In Part II (following article), we consider deep-inelastic scattering on the polarized deuteron with spectator
nucleon tagging as a special case of target fragmentation. The semi-inclusive structure functions are computed
by separating nuclear and hadronic structure, and the polarization observables are explored
as functions of the tagged nucleon momentum.
\end{abstract}
\preprint{JLAB-THY-26-4663}
\maketitle
\tableofcontents
\section{Introduction}
\label{sec:intro}
This is Part I of a study of semi-inclusive deep-inelastic scattering on a polarized spin-1 target.
It presents the general form of the semi-inclusive cross section and polarization
observables for the spin-1 target. {\secondpart} considers deep-inelastic scattering on the
polarized deuteron with spectator nucleon tagging.

Deep-inelastic lepton scattering (DIS) is a principal tool for exploring the structure of hadrons
and nuclei and the dynamics of strong interactions at short distances.
The momentum transfer $Q^2 \gtrsim 1$ GeV$^2$ resolves structures at transverse distances
much smaller than the hadronic size $\sim 1$ fm.
The energy transfer $\nu \gg 1$ GeV causes hadron production over a wide rapidity interval
and separates current and target fragments in the final state.
Basic information comes from measurements of the lepton-target differential cross section without
specification of the hadronic final state (inclusive DIS).
Further information is obtained from measurements of the lepton and target spin dependence
(polarized DIS).
Even more information becomes available with measurements of single-identified hadrons in the
final state (semi-inclusive DIS or SIDIS, unpolarized and polarized).
Combinations of these measurements make it possible to probe specific elements of
the target structure or the final state.

Most theoretical and experimental efforts in polarized DIS have focused on the spin-1/2
target of the nucleon, see reviews~\cite{Lampe:1998eu,Kuhn:2008sy,Aidala:2012mv,Deur:2018roz}
and references therein. Measurements of polarized inclusive DIS on the nucleon probe the
polarized quark and gluon distributions in the context of the QCD factorization of the
cross section \cite{Collins:1989gx,Kodaira:1998jn,Collins:2011zzd}.
Measurements of polarized SIDIS on the nucleon in the
current fragmentation region provide information on the flavor decomposition of the polarized quark
distributions and spin-orbit effects in the transverse momentum dependent (TMD)
distribution and fragmentation functions \cite{Boussarie:2023izj}.
While several experiments in polarized DIS and
SIDIS \cite{SpinMuon:1997yns,HERMES:2004zsh,COMPASS:2009kiy,COMPASS:2008isr,COMPASS:2023vhr}
have used the spin-1 deuteron as a target, they have been interpreted in terms of scattering
on a polarized neutron (after subtracting the proton), with the role of the deuteron
limited to providing the effective polarization of the neutron.

Fewer efforts have been devoted to studying the specific polarization effects arising
in DIS on targets with spin $>$1/2. A new aspect of spin-1 targets is the possibility of
tensor polarization in addition to vector polarization, giving rise to new structures
in the cross section \cite{Frankfurt:1983qs,Hoodbhoy:1988am}.
Inclusive DIS on the tensor-polarized deuteron was explored
at HERMES \cite{HERMES:2005pon} and will be studied in dedicated experiments
at JLab \cite{Long:2014fda,Slifer:2013vma,Cosyn:2024drt,Poudel:2025nof}.
The theoretical interpretation
focuses on nuclear binding effects on the partonic structure at large $x$ ($\gtrsim$ 0.1)
\cite{Frankfurt:1983qs,Khan:1991qk,Cosyn:2017fbo,Miller:2013hla},
and on nuclear shadowing dynamics at small $x$ ($\ll$ 0.1)
\cite{Bora:1997pi,Edelmann:1997ik,Edelmann:1997qe,Frankfurt:2006am}.
Recent work has explored SIDIS on spin-1 targets in the current fragmentation region
and the new TMD distributions arising in this
context \cite{Bacchetta:2000jk,Boer:2016xqr,Kumano:2024fpr,Zhao:2025vol,Kumano:2025rai}.
The tensor polarization gives rise to new spin-orbit effects in SIDIS observables and
TMD distributions, making it possible to explore new aspects of partonic structure.

DIS on the deuteron with detection of a ``slow'' nucleon (with momenta $\sim$ few 100 MeV
in the nuclear rest frame) is a special case of SIDIS in the target fragmentation region.
In the context of QCD factorization, target fragmentation measurements are described
by fracture functions, which combine aspects of parton distribution and fragmentation functions
and express both target structure and hadronization dynamics
\cite{Trentadue:1993ka,Collins:1997sr,Anselmino:2011ss}.
In the case of slow nucleon production in deuteron DIS, the dominant production mechanism
is the nuclear breakup, and the cross section can be computed on this basis (``spectator nucleon tagging'')
\cite{Frankfurt:1983qs,Frankfurt:1981mk};
this approximation is compatible with QCD factorization and amounts to a particular nonperturbative
model of the fracture functions \cite{Strikman:2017koc}. It presents a unique situation where the SIDIS
cross section and spin observables can be predicted on the basis of well-known nuclear dynamics, formulated
in nucleon degrees of freedom and their interactions. In particular, the tensor-polarized structures
and spin-orbit effects unique to spin-1 SIDIS could be explored in spectator nucleon tagging
with the polarized deuteron \cite{Frankfurt:1983qs,Cosyn:2020kwu}.

Measurements of DIS on the deuteron with spectator nucleon tagging are performed in fixed-target scattering
experiments at Jefferson Lab, using specialized detectors for slow protons or neutrons
emerging from the target (BONuS, ALERT, Deeps, BAND, LAD)
\cite{CLAS:2011qvj,CLAS:2014jvt,Armstrong:2017wfw,Albayrak:2024vcy,CLAS:2005ekq,EMCSRC:2011,Segarra:2020txy}.
Measurements of spectator tagging are also planned at the Electron-Ion Collider (EIC), using the
far-forward detectors for charged and neutral beam fragments \cite{Jentsch:2021qdp,AbdulKhalek:2021gbh}.
The collider setup offers several advantages for nuclear breakup detection (see \secondpart\ for details).
Such measurements could also be performed on the polarized deuteron if polarized deuteron
beams become available at EIC; see Refs.~\cite{Huang:2020uui,Huang:2021gbi,Huang:2025gqx}
for the technical prospects.

In this work we develop the theoretical framework for SIDIS on a polarized spin-1 target
and apply it to scattering on the polarized deuteron with spectator nucleon tagging.
The presentation consists of two parts.

In Part I (this article), we derive the general form of the semi-inclusive cross section and polarization
observables for the spin-1 target. A relativistically covariant formulation in terms of 4-vectors
and invariant polarization parameters is employed. The target polarization is described by a covariant
spin density matrix with vector and tensor polarization. The spin and azimuthal angle dependence of the
semi-inclusive cross section is derived and parametrized in terms of invariant structure functions.
The expressions obtained in this part are general (no assumptions about particle production dynamics)
and valid in all regions of the deep-inelastic final state (current and target fragmentation regions).
They can be used for any kind of SIDIS measurement on a polarized spin-1 target.

In Part II (following article, Ref.~\cite{CW_kin_deut}),
we consider the specific case of DIS on the polarized deuteron with spectator nucleon tagging.
The semi-inclusive structure functions are computed by separating nuclear and hadronic structure,
and the polarization observables are explored as functions of the tagged nucleon momentum.
The findings are interpreted using the paradigm of ``selection of nuclear configurations''
by the tagged nucleon momentum.
The results obtained in this part involve approximations in the separation of nuclear and hadronic
structure (light-front quantization, impulse approximation) and empirical parametrizations
of the low-energy nuclear structure. They can be used for quantitative analysis and simulations
of experiments in DIS on the deuteron with spectator nucleon tagging.

The derivation of the SIDIS cross section for the spin-1 target here follows earlier work
for the spin-1/2 target \cite{Kotzinian:1994dv,Diehl:2005pc,Bacchetta:2006tn,Bastami:2018xqd,Donnelly:2023rej}.
In the spin-1/2 case the target polarization is characterized by three parameters, and the semi-inclusive
cross section involves 18 independent structures with a distinctive dependence on the beam and target
polarization and the azimuthal angle of the identified hadron. In the spin-1 case, the target
polarization involves five additional tensor polarization parameters, and the semi-inclusive cross
section exhibits 23 additional structures, some with new azimuthal dependences not found in the spin-1/2 case.
When integrating over the identified hadron momentum, four of these semi-inclusive tensor-polarized structures
reduce to the well-known inclusive tensor-polarized structures (denoted by $b_1$ to $b_4$)
\cite{Hoodbhoy:1988am}; the others integrate to zero.
The spin-1 SIDIS cross section thus has a much richer structure than the spin-1/2 one,
providing opportunities to study spin-orbit phenomena and structures not present in the spin-1/2 system.
The form of the spin-1 SIDIS cross section was analyzed recently in Ref.~\cite{Zhao:2025vol}.
It can also be compared with the form of the exclusive deuteron electrodisintegration
cross section obtained in earlier nuclear physics studies \cite{Leidemann:1991qs}
(further references are given in {\secondpart}).

The definition of the azimuthal angle variables and coordinate system is an important issue in SIDIS phenomenology.
For the azimuthal angles of the target polarization and the final-state hadron momentum we follow
the Trento convention (see below) \cite{Bacchetta:2004jz}. For the coordinate system we choose the
direction of the $z$-axis opposite to the virtual photon momentum, along the direction of the target
momentum in the center-of-mass frame, which simplifies the construction of target spin states
and the use of light-front quantization for target structure, see {\secondpart}. The correspondence
with other conventions can easily be established.

The article is organized as follows.
Section~\ref{sec:kin} describes the general setup, including kinematic variables,
basis 4-vectors, collinear frames, and final-state variables.
Section~\ref{sec:densitymatrix} presents the description of the polarized spin-1 target,
including the covariant spin density matrix, invariant polarization parameters,
and the tensor and spin quantum number representations of the density matrix.
Section~\ref{sec:cross_sec} presents the derivation of the differential cross section,
including the hadronic tensor, its decomposition and structure functions,
the final form of the spin-1 semi-inclusive cross section including the azimuthal angle
and spin dependence (vector and tensor polarization), and the correspondence with
the spin-1 inclusive cross section.
Section~\ref{sec:observables} describes the spin observables in spin-1 SIDIS,
including the preparation of target polarization using combinations of
pure spin states, the separation of vector and tensor polarized structures,
azimuthal harmonics, and spin asymmetries.
Section~\ref{sec:summary} presents a summary and possible extensions
of the present work.

Appendix~\ref{app:correspondence} summarizes the changes in definitions
relative to the earlier work in Ref.~\cite{Cosyn:2020kwu}.
Appendix~\ref{app:density_matrix} describes an alternative derivation of the
spin-1 spin density matrix using quantum-mechanical operators.
Appendix~\ref{sec:helicity_amps} discusses the representation of the spin-1
semi-inclusive structure functions as helicity amplitudes, which is useful
for validating the number of types of independent structures.
Appendix~\ref{app:higher_spin} outlines the extension of the
structural decomposition of the cross section to targets of higher spin $(>1)$.

\section{Process and variables}
\label{sec:kin}
\subsection{Kinematic variables}
We consider the scattering of a polarized lepton $e$ on a polarized spin-1 target $A$,
with detection of the scattered lepton $e'$ and a specified hadron $h$ in the final state,
and no restriction on the rest of the hadronic final state $X$ (semi-inclusive scattering),
\begin{align}
e(l| \lambda_e) + A(P| S, T) \rightarrow e'(l') + X + h(P_h).
\label{process}
\end{align}
$l$ and $l'$ are the initial and final lepton 4-momenta; we neglect the lepton mass.
$P$ is the target 4-momentum, $P^2 = M^2$, and $P_h$ is the detected hadron 4-momentum, $P_h^2 = M_h^2$;
we follow the notation of Ref.~\cite{Bacchetta:2006tn} for the spin-1/2 target.
The initial state is characterized by the squared center-of-mass energy, described by the invariant variable
\begin{align}
s \equiv (l + P)^2.
\end{align}
The four-momentum transfer to the target is
\begin{align}
&q \equiv l - l', \hspace{1em} Q^2 \equiv -q^2 .
\end{align}  
The hadronic scattering process is characterized by the energy and momentum transfer
to the target, described by the invariant scaling variables\footnote{$x_A$ 
denotes the Bjorken scaling variable for the target $A$ with $0 < x_A < 1$.
In the applications to the deuteron ($A = D$) in {\secondpart}, we shall also use
the effective scaling variable for the nucleon in the deuteron, which is
defined as $x = 2 x_D$ and takes values $0 < x < 2$.}
\begin{align}
&x_A =\frac{Q^2}{2 (P q)},
\hspace{1em}
y=\frac{(Pq)}{(Pl)},
\label{scaling_variables_general}
\end{align}
which satisfy $0 < x_A \leq 1$ and $0 < y < 1$. The kinematic boundaries in a
given setup are determined by the condition
\begin{align}
Q^2 &= x_A y (s - M^2) .
\end{align}
The final state with the identified hadron $h$ is characterized by three additional
momentum variables, which are defined using a specific reference frame (see below).

The initial lepton in the scattering process Eq.~(\ref{process}) is in a pure spin state
described by the helicity $\lambda_e = \pm 1/2$. The spin-1 target $A$ is in a mixed spin state
described by a density matrix, parametrized by the vector and tensor polarization parameters, $S$ and $T$;
its definition and properties are described in Sec.~\ref{sec:densitymatrix}.

\subsection{Basis vectors}
\label{subsec:basis_vectors}
For describing the SIDIS process in a relativistically covariant manner it is useful
to introduce a basis of orthonormal 4-vectors derived from the particle momenta.
One distinguishes the ``longitudinal'' subspace spanned by vectors $q$ and $P$, and the
``transverse'' subspace orthogonal to it.

In the longitudinal subspace we introduce two different sets of orthonormal vectors,
adapted to different purposes as explained in the following. Set I consists of the vectors 
\begin{subequations}
\label{basis_longit_set1}
\begin{align}
&e_q \equiv \frac{q}{\sqrt{-q^2}},
\hspace{1em}
e_L \equiv \frac{P + (e_q P) e_q}{\sqrt{P^2 + (e_q P)^2}},
\\[1ex]
&e_q^2 = -1, \hspace{1em} e_L^2 = 1, \hspace{1em} (e_q e_L) = 0. 
\end{align}
\end{subequations}
Set II consists of the vectors
\begin{subequations}
\label{basis_longit_set2}
\begin{align}
&e_P \equiv \frac{P}{\sqrt{P^2}},
\hspace{1em}
e_{L\ast} \equiv \frac{q - (e_P q) e_P}{\sqrt{-q^2 + (e_P q)^2}},
\\[1ex]
& e_P^2 = 1, \hspace{1em} e_{L\ast}^2 = -1, \hspace{1em} (e_P e_{L\ast}) = 0.
\end{align}
\end{subequations}
The two sets are related by a hyperbolic rotation (equivalent to a Lorentz transformation),
\begin{subequations}
\label{basis_set1_set_2}
\begin{align}
&e_L = \cosh\alpha\, e_P + \sinh\alpha \, e_{L\ast},
\\[1ex]
&e_q = \sinh\alpha\, e_P + \cosh\alpha \, e_{L\ast},
\\[1ex]
&\sinh\alpha = 1/\gamma,
\hspace{1em}
\cosh\alpha =  \sqrt{1 + \gamma^2} / \gamma,
\end{align}
\end{subequations}
where
\begin{align}
\gamma \equiv \frac{\sqrt{-q^2 P^2}}{(Pq)} = \frac{2Mx_A}{Q} .
\label{gamma_def}
\end{align}
Both sets are complete in the longitudinal subspace,
\begin{align}
e_L^\mu e_L^\nu - e_q^\mu e_q^\nu =
e_P^\mu e_P^\nu - e_{L\ast}^\mu e_{L\ast}^\nu
\equiv g_L^{\mu\nu}.
\end{align}

The projector on the transverse subspace is 
\begin{align}
g_T^{\mu\nu}
&\equiv g^{\mu\nu} - g_L^{\mu\nu},
\hspace{1em}
g^{\mu\nu} = g_L^{\mu\nu} + g_T^{\mu\nu}.
\end{align}
The transverse component of a 4-vector $V^\mu$ is denoted as
\begin{align}
&V_T^\mu \equiv g_T^{\mu\nu} V_\nu,
\hspace{1em}
V_T^2 < 0;
\label{transverse_component}
\end{align}
it is a spacelike 4-vector, subject to the Lorentz metric.
The antisymmetric tensor in the transverse subspace is 
\begin{align}
\epsilon_T^{\mu\nu} \equiv \epsilon^{\mu\nu\rho\sigma} e_{L,\rho} e_{q,\sigma}
= &\epsilon^{\mu\nu\rho\sigma} e_{P,\rho} e_{L\ast,\sigma},
\hspace{1em}
\epsilon^{0123} \equiv 1.
\end{align}

%
%
\begin{figure*}[t]
\begin{center}
\includegraphics[width=0.7\textwidth]{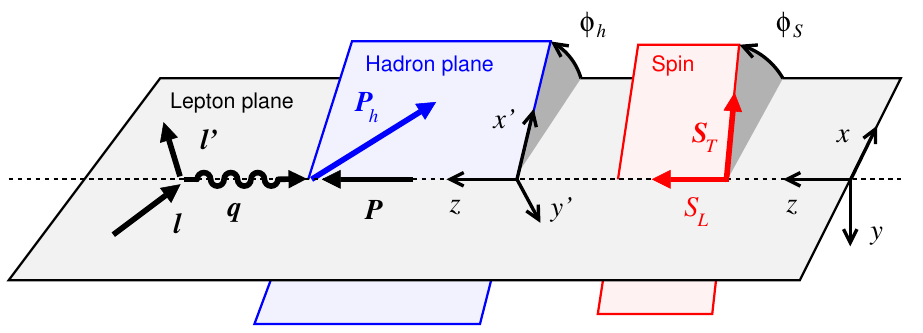}
\end{center}
\caption{
\label{fig:kin}
Azimuthal angles and coordinate systems in the collinear frames.
The angles $\phi_h$ and $\phi_S$ conform to the Trento convention \cite{Bacchetta:2004jz}.
The $xyz$ coordinate system is aligned with the lepton plane, with the $z$ direction opposite
to $\bm{q}$ and the $x$ direction along $\bm{l}_T = \bm{l}_T'$. The $x'y'z$ coordinate system
is aligned with the hadron plane, with the $x'$ direction along $\bm{P}_{hT}$.}
\end{figure*}
In the transverse subspace the basis vectors can be aligned either
with the lepton momentum $l$ or with final hadron momentum $P_h$.
We define basis vectors aligned with the lepton momentum as
\begin{align}
&e_x \equiv \frac{l_T}{\sqrt{-l_T^2}},
\hspace{1em}
e_y^\mu \equiv \epsilon_T^{\mu\nu} e_{x, \nu},
\hspace{1em}
e_{x}^2, e_{y}^2 = -1.
\label{basis_transv_lepton}
\end{align}
$e_x$ is along the lepton transverse momentum, and $e_y$ is normal to it.
Note that $e_x$ is a vector, and $e_y$ is a pseudovector.
In a similar way we define basis vectors aligned with the final hadron momentum as
\begin{align}
&e_{x'} \equiv \frac{P_{hT}}{\sqrt{-P_{hT}^2}},
\hspace{1em}
e_{y'}^\mu \equiv \epsilon_T^{\mu\nu} e_{x', \nu},
\hspace{1em}
e_{x',y'}^2 = -1.
\label{basis_transv_hadron}
\end{align}
The two sets of transverse 4-vectors are related by a planar rotation
\begin{subequations}
\label{basis_transv_angle}
\begin{align}
e_{x'} &= \cos\phi_h e_x - \sin\phi_h e_y,
\\
e_{y'} &= \sin\phi_h e_x + \cos\phi_h e_y.
\end{align}
\end{subequations}
The angle $\phi_h$ can be expressed through the scalar products of the 4-vectors,
\begin{align}
\cos\phi_h = -(e_x e_{x'}),
\hspace{1em}
\sin\phi_h = (e_y e_{x'}),
\label{basis_transv_angle_scalar}
\end{align}
and is thus defined in a relativistically invariant manner and without reference to a coordinate system.
The definition of the angle $\phi_h$ here follows the Trento convention \cite{Bacchetta:2004jz}
(see further discussion in Sec.~\ref{subsec:collinear_frames}).
Note that the definition of the angle differs from the one in the earlier work
of Ref.~\cite{Cosyn:2020kwu} (see Appendix~\ref{app:correspondence} for reference).
Both sets of basis vectors are complete in the transverse subspace
\begin{align}
- e_{x}^\mu e_{x}^\nu - e_{y}^\mu e_{y}^\nu
= - e_{x'}^\mu e_{x'}^\nu - e_{y'}^\mu e_{y'}^\nu
= g_T^{\mu\nu} .
\end{align}

In summary, each of the 4-vector sets
\begin{align}
\{ e_L, e_q, e_x, e_y \},
\hspace{1em}
\{ e_P, e_{L\ast}, e_x, e_y \}
\end{align}
(or the equivalent sets with $e_{x'}, e_{y'}$) forms a natural basis of 4-dimensional spacetime
and can be used to expand the vectors and tensors characterizing the scattering process;
depending on the quantity, one or the other sets will be more convenient.
The basis vectors are defined without reference to any specific frame or coordinate system
and can be used accordingly. A simple geometric interpretation of the basis vectors
becomes possible in the collinear frames described in the following subsection. 

\subsection{Collinear frames}
\label{subsec:collinear_frames}
The SIDIS process is naturally described in a so-called collinear frame, defined by the
condition that the target momentum and the momentum transfer be collinear 3-vectors,
$\bm{P}\parallel\bm{q}$ (see Fig.~\ref{fig:kin}). The condition defines not a single
frame but a class of frames, whose members are related by boosts along the collinear axis.
It contains several frames of special interest: the target rest frame ($\bm{P} = 0$);
the photon-target center-of-mass frame ($\bm{P} + \bm{q} = 0$), and a frame
with zero energy transfer ($q^0 = 0$, Breit frame). The following discussion
applies to any frame in the class and does not assume a specific member unless
specified otherwise.

The collinear axis is chosen as the $z$-axis, with $\bm q$ pointing in the negative
$z$-direction, $\bm{q} = q^z \bm{e}_z$ with $q^z < 0$.
With this choice any frame reached by a boost along the $z$-axis will
also have $q^z < 0$, because the sign of $q^z$ remains unchanged under boosts.
This is because the 4-vector $q$ is spacelike, $q^2 < 0$, and there is no frame
where it can have $q^z = 0$, which would imply that only the time component is nonzero
and the 4-vector is timelike. The value of $q^z < 0$ serves as a boost parameter
selecting a particular member of the class of collinear frames (other ways to select
a frame based on light-front components are discussed in {\secondpart}).
The target momentum can have components $P^z > 0$ or $P^z < 0$, depending on the particular frame.

The components of the
longitudinal basis 4-vectors of Set I, Eq.~(\ref{basis_longit_set1}), in the collinear frame
can be expressed as
\begin{align}
& e_q = Q^{-1} (\omega, q^z \, \bm{e}_z),
\hspace{1em}
e_L = Q^{-1} (-q^z, -\omega \, \bm{e}_z ),
\nonumber \\[1ex]
& \omega \equiv \sqrt{(q^z)^2 - Q^2}, \hspace{1em} |q^z| > Q, \hspace{1em} q^z < 0.
\end{align}
The components of Set II can be obtained from Eq.~(\ref{basis_set1_set_2}).

The transverse component of the lepton momentum, together with the collinear axis,
defines the lepton plane (see Fig.~\ref{fig:kin}). We denote the coordinate system aligned with
the lepton plane by $xyz$, with $x$ along the initial lepton transverse momentum (which is equal
to the final transverse lepton momentum, because the momentum transfer is in the longitudinal direction),
and $y$ along the normal direction, such that $xyz$ forms a right-handed system.
In this coordinate system the basis 4-vectors Eq.~(\ref{basis_transv_lepton}) have components
\begin{align}
&e_{x} = (0, \bm{e}_{x}),
\hspace{2em}
e_{y} = (0, \bm{e}_{y});
\label{collinear_frame_lepton}
\end{align}
their spatial components coincide with the 3-dimensional unit vectors along the directions.
The transverse component of the final hadron momentum, together with the collinear axis,
defines the hadron plane (see Fig.~\ref{fig:kin}). We denote the coordinate system aligned with
the hadron plane by $x'y'z$, with $x'$ along the hadron transverse momentum,
and $y'$ the right-handed normal. The basis 4-vectors Eq.~(\ref{basis_transv_hadron}) have components
\begin{align}
&e_{x'} = (0, \bm{e}_{x'}),
\hspace{2em}
e_{y'} = (0, \bm{e}_{y'}).
\label{collinear_frame_hadron}
\end{align}
Equations~(\ref{collinear_frame_lepton}) and (\ref{collinear_frame_hadron}) describe the
transverse basis vectors in any collinear frame; their form is not affected by boosts
along the $z$-axis.

The angle $\phi_h$ in Eq.~(\ref{basis_transv_angle}) becomes the angle
between the lepton and the hadron planes in the collinear frame (see Fig.~\ref{fig:kin}). It is defined according
to the Trento convention \cite{Bacchetta:2004jz} and takes values $\phi_h \in [0, 2 \pi]$.
In our coordinate systems $\phi_h$ is the angle of the rotation taking the $xy$- to the $x'y'$-axes,
measured in the negative mathematical sense (rotation of the positive $x$-axis away from the
positive $y$-axis). Hadron momenta with $P_h^y < 0$ ($> 0$) have azimuthal angles $\phi_h < \pi$ ($>\pi$).
Note that the angle is unambiguously defined in terms of the particle 4-momenta by
Eq.~(\ref{basis_transv_angle_scalar}), which does not refer to a coordinate system.
Reference~\cite{Bacchetta:2004jz} and other works use a coordinate system where the
$z$-axis is along $\bm{q}$, which differs from ours by reversing the direction of the
$z$- and $y$-axes (see Fig.~\ref{fig:kin}); the expressions presented in the following
could easily be transcribed to that coordinate system without changing the definition of $\phi_h$.

A special collinear frame is the target rest frame, where $P = (M, \bm{0})$.
In this frame the longitudinal basis 4-vectors of Set II Eq.~(\ref{basis_longit_set2}) have components
\begin{align}
e_P = (1, \bm{0}), \hspace{1em} e_{L\ast} = (0, -\bm{e}_{z}).
\label{set2_collinear_frame}
\end{align}
The spatial component of $e_{L\ast}$ is the negative of the unit vector in the $z$ direction.
When using the basis 4-vectors of Set II (together with the transverse basis vectors) to expand
vectors and tensors describing the scattering process, one can make a direct connection with
the corresponding quantities in the target rest frame. In Sec.~\ref{sec:densitymatrix} we use
the basis $\{ e_P, e_{L\ast}, e_{x'}, e_{y'} \}$ to expand the target polarization
vector and tensor and connect it with the rest-frame polarization.

It should be noted that in actual experiments the relation between the collinear frame
and the lab frame depends on the final-state lepton and hadron momenta, which enter
in the definition of the basis vectors. Even for fixed final lepton momentum (i.e., for
fixed $x_A$ and $Q^2$) the orientation of the transverse vectors depends on the
observed hadron momentum.

\subsection{Final-state hadron variables}
\label{subsec:final_state}
The momentum of the final-state hadron in the SIDIS process is characterized by three kinematic variables.
They refer to the hadron's longitudinal and transverse momentum in the collinear frame and can be
defined in terms of invariants. The hadron transverse momentum is defined invariantly
as the norm of the 4-momentum projection on the transverse subspace, Eq.~(\ref{transverse_component}),
\begin{equation}
|\bm P_{hT}| \equiv {\textstyle\sqrt{-P_{hT}^2} = \sqrt{-(g_T P_{h})^2}};
\label{P_hT_invariant}
\end{equation}
its angle $\phi_h$ is defined invariantly through the scalar products with the basis vectors
aligned with the lepton momentum, Eq.~(\ref{basis_transv_angle_scalar}).
In the collinear frame, $|\bm P_{hT}|$ and $\phi_h$ describe the magnitude and
azimuthal angle of the transverse momentum as specified in Fig.~\ref{fig:kin}.

To characterize the hadron longitudinal momentum several variables have been introduced,
motivated by various physical considerations.
One invariant variable is the ratio
\begin{equation}
z_h \equiv \frac{(PP_h)}{(Pq)} ,
\label{z_h_def}
\end{equation}
which can be interpreted as the fraction of the virtual photon energy carried by hadron in the
target rest frame; this variable is commonly used in the analysis of SIDIS in the
current fragmentation region, where $z_h = \mathcal{O}(1)$. Another invariant variable is
\begin{align}
\zeta_h \equiv \frac{(nP_h)}{(nP)}, \hspace{1em} n \equiv e_P + e_{L\ast},
\hspace{1em} n^2 = 0,
\label{zeta_h_def}
\end{align}
which is the fraction of the light-cone momentum of the target carried by the produced hadron;
this variable is used in the analysis of target fragmentation and nuclear breakup processes
(a version rescaled by the nuclear mass number is used in {\secondpart}).
In the collinear frame, $\zeta_h = P_h^+ / P^+$, where
$P^+ \equiv P^0 + P^z$ is the light-cone plus momentum.
A related variable is the rapidity in the collinear frame
\begin{align}
y \equiv \frac{1}{2} \ln \frac{P_h^0 + P_h^z}{P_h^0 - P_h^z},
\end{align}
which changes by a constant under longitudinal boosts (the origin of the scale depends
on the boost parameter of the frame). For other choices of the longitudinal
momentum variable and their interpretation, see Ref.~\cite{Strikman:2017koc}.

Figure~\ref{fig:kin} shows a situation where the longitudinal component of the
final hadron momentum is along the photon direction, $P_h^z < 0$, which is typically
the case in current fragmentation. The longitudinal and transverse momentum variables
introduced here also apply to situations where final hadron momentum is along the
target momentum in the photon-target center-of-mass frame, which is typically the
case in target fragmentation. In {\secondpart} we consider detection of the
spectator nucleon in DIS on the deuteron, which is an instance of target fragmentation.
\section{Polarized spin-1 target}
\label{sec:densitymatrix}
\subsection{Covariant spin-1 density matrix}
\label{subsec:covariant_density_matrix}
The mixed spin state of the spin-1 target in the scattering process Eq.~(\ref{process})
is described by a density matrix. We now determine the general form of this matrix
for a moving target, its parametrization through invariant polarization parameters,
and the expansion of the density matrix in the basis 4-vectors of Sec.~\ref{sec:kin}.

Pure spin states of a spin-1 particle with mass $M$ and 4-momentum $P$ are described
covariantly by a 4-vector wave function $\epsilon^\alpha (\lambda)$, with
\begin{subequations}
\begin{align}
& P^\alpha \epsilon_\alpha (\lambda) \; = \; 0,
\\[2ex]
& \epsilon^{\alpha\ast} (\lambda) \epsilon_\alpha (\lambda') \; = \; - \delta(\lambda, \lambda') ,
\\
& \sum_\lambda \epsilon^\alpha (\lambda) \epsilon^{\beta\ast} (\lambda) \; = \; - g^{\alpha\beta} +
\frac{P^\alpha P^\beta}{M^2}.
\end{align}
\end{subequations}
$\lambda = (+1, 0, -1)$ denotes the spin quantum number. Here and in the following, it is assumed that the wave
function of the moving particle is obtained from that of the particle at rest by a boost,
\begin{align}
\epsilon^\alpha (\lambda | \bm{P} = 0)
\rightarrow \epsilon^\alpha (\lambda | \bm{P}),
\label{boost}
\end{align}
and $\lambda$ is understood as the spin projection in the particle's rest frame with respect to a
given axis. The specific form of the moving wave function depends on the choice of the boost;
e.g. canonical boosts (along the direction of $\bm{P}$) or light-front boosts (a sequence of boosts along the
longitudinal and transverse directions). The following considerations apply to any choice of the boost.

A mixed spin state of the spin-1 particle in the rest frame is described by a density matrix
$\rho(\lambda, \lambda')$, a hermitean $3 \times 3$ matrix with unit trace,
\begin{subequations}
\begin{align}
&\rho(\lambda, \lambda') = \rho^\ast(\lambda', \lambda) ,
\\[1ex]
& \sum_{\lambda} \rho(\lambda, \lambda) = 1 .
\end{align}
\end{subequations}
In analogy with the description of pure states, the mixed state of the moving particle is then described
covariantly by a 4-tensor density matrix
\begin{align}
\rho^{\alpha\beta} \equiv
\sum\limits_{\lambda,\lambda'} \rho(\lambda,\lambda')\epsilon^\alpha(\lambda)\epsilon^{*\beta}(\lambda'),
\label{density_tensor}
\end{align}
with 
\begin{subequations}
\label{density_tensor_conditions}
\begin{align}
& P_\alpha \rho^{\alpha\beta} \; = \; 0, \hspace{1em} \rho^{\alpha\beta} P_\beta \; = \; 0,
\\[1ex]
& \rho^{\alpha\beta} = \rho^{\beta\alpha\ast} ,
\\[1ex]
& \rho^{\alpha}_{\;\;\alpha} \; = \; -1 .
\end{align}
\end{subequations}
A spin observable characterizing the spin-1 system is given by a matrix in the spin quantum numbers,
$O(\lambda', \lambda)$, which may be of 4-vector or tensor character and carry additional indices not shown here.
In the 4-tensor representation it corresponds to a bilinear form
\begin{align}
O(\lambda', \lambda) \; \equiv \;  \epsilon^{\beta\ast}(\lambda') O_{\beta\alpha} \epsilon^\alpha (\lambda) ,
\end{align}
where $O_{\beta\alpha}$ is a tensor in the indices $\alpha, \beta$. The expectation value in the mixed state
is computed as
\begin{align}
\langle O \rangle \; \equiv \; \sum_{\lambda, \lambda'} \rho(\lambda, \lambda') 
O(\lambda', \lambda)
\; = \; \rho^{\alpha\beta} O_{\beta\alpha} .
\label{observable_average}
\end{align}
As a special case, the 3-dimensional matrix elements $\rho(\lambda, \lambda')$
are obtained by choosing $O_{\beta\alpha}$ as the tensor product of the corresponding 4-vectors,
\begin{align}
\rho(\lambda, \lambda') =  \rho^{\alpha\beta}
\epsilon_\beta (\lambda') \epsilon^\ast_\alpha (\lambda).
\label{projection}
\end{align}
This formula can be used to evaluate the elements of the rest-frame density matrix in any frame
where the particle is moving, if the explicit form of the 4-vector wave functions is known.

The 4-tensor density matrix Eq.~(\ref{density_tensor}) can be parametrized covariantly as 
\begin{align}
\rho^{\alpha\beta} &= \frac{1}{3}\left(-g^{\alpha\beta}+\frac{P^\alpha P^\beta}{M^2} \right)
+ \frac{i}{2M}\epsilon^{\alpha\beta\gamma\delta}P_\gamma S_\delta - T^{\alpha\beta}.
\label{density_tensor_parametrization}
\end{align}
The terms are known as the unpolarized, vector-polarized, and tensor-polarized term.
The parameters are the real axial 4-vector $S^\alpha$, 
\begin{align}
P^\alpha S_\alpha = 0,
\label{vector_condition}
\end{align}
and the real symmetric traceless 4-tensor $T^{\alpha\beta}$,
\begin{subequations}
\label{tensor_conditions}
\begin{align}
& P_\alpha T^{\alpha\beta} \; = \; 0, \hspace{1em}  \; T^{\alpha\beta}  P_\beta \; = \; 0,
\\[1ex]
& T^{\alpha\beta} = T^{\beta\alpha} ,
\\[1ex]
& T^{\alpha}_{\;\;\alpha} \; = \; 0 .
\end{align}
\end{subequations}
The form of Eq.~(\ref{density_tensor_parametrization}) follows from the conditions
of Eqs.~(\ref{density_tensor_conditions}). It applies to any choice of boost defining
the moving spin states, Eq.~(\ref{boost}); the differences between choices are encoded
in the values of the parameters $S^\alpha$ and $T^{\alpha\beta}$. Thus the polarization
of the moving spin-1 particle is completely specified by the covariant parameters
$S^\alpha$ and $T^{\alpha\beta}$ and can be evaluated in any frame.

\subsection{Invariant polarization parameters}
\label{subsec:invariant_polarization}
For describing the scattering process Eq.~(\ref{process}) we expand the parameters
of Eq.~(\ref{density_tensor_parametrization}) in the basis 4-vectors of Sec.~\ref{sec:kin}.
Since $S^\alpha$ and $ T^{\alpha\beta}$ are orthogonal to the target 4-momentum $P^\mu$,
it is convenient to expand them in the longitudinal vectors of Set II, Eq.~(\ref{basis_longit_set2}).
For the transverse vectors we use the ones aligned with the final-state hadron momentum,
Eq.~(\ref{basis_transv_hadron}). We define the expansion coefficients as
\begin{subequations}
\label{eq:poltensor_LT}
\begin{align}
&(e_{L\ast}S) \equiv S_L,
\\[1ex]
&-( e_{x'} S) \equiv S_T 
\cos(\phi_S - \phi_h),
\\[1ex]
&-( e_{y'} S) \equiv -S_T 
\sin(\phi_S - \phi_h),
\\[1ex]
&(e_{L\ast} Te_{L\ast}) \equiv  T_{LL},
\\[1ex]
&-(e_{L\ast} T e_{x'}) \equiv 
 T_{LT}\cos(\phi_{T_L} - \phi_h),
\\[1ex]
&-(e_{L\ast} T e_{y'}) \equiv 
-T_{LT}\sin(\phi_{T_L} - \phi_h),
\\[1ex]
&( e_{x'} T e_{x'})-( e_{y'} T e_{y'}) \equiv 
 T_{TT}\cos(2\phi_{T_T} - 2\phi_h),
\\[1ex]
&( e_{x'} T e_{y'}) \equiv 
-\tfrac{1}{2} T_{TT}\sin(2\phi_{T_T} - 2\phi_h).
\end{align}
\end{subequations}
The vector polarization is now expressed by the three parameters 
$\{ S_L, S_T, \phi_S \}$, and the tensor polarization by the five
parameters $\{ T_{LL}, T_{LT}, T_{TT}, \phi_{T_L} , \phi_{T_T} \}$.
Notice that we do not introduce an additional parameter for 
$( e_{x'} T e_{x'})+( e_{y'} T e_{y'})$, as this combination is equal to $-(e_{L\ast} Te_{L\ast})$ 
due to the zero trace of the tensor $T$, Eq.~(\ref{tensor_conditions}).
The parameters are defined in terms of scalar products of 4-vectors and are thus invariant.
We refer to them as the invariant polarization parameters.
The parameters $S_T, T_{LT}, T_{TT}$ are $\geq 0$ and can be interpreted as intensities;
this can be shown by evaluating the invariant parameters in the target rest frame.

In Eq.~(\ref{eq:poltensor_LT}) the polarization 4-vector and tensor are contracted
with the basis vector $e_{L\ast}$ of Set II. Because of the conditions Eq.~(\ref{vector_condition})
and (\ref{tensor_conditions}), these contractions can equivalently be computed with the
basis vector $e_q$ of Set I, related to $e_{L\ast}$ by Eq.~(\ref{basis_set1_set_2}),
\begin{align}
&S_L = (e_{L\ast}S) = \frac{\gamma}{\sqrt{1+\gamma^2}} \, (e_q S),
\end{align}
and similarly for the contractions with $T$. These relations
will be used when evaluating the polarized cross section in Sec.~\ref{subsec:structure_functions}.

The invariant polarization parameters include scalars with even and odd parity. The parity can be inferred from the
parity of the vectors and tensors entering in the scalar products. One finds
\begin{subequations}
\label{polarizations_parity}
\begin{align}
\textrm{even:} \hspace{1em} & (e_{y'} S ), (e_{L\ast}Te_{L\ast}), (e_{L\ast}Te_{x'}),
\nonumber \\
& (e_{x'}Te_{x'}) - (e_{y'}Te_{y'}),
\\[1ex]
\textrm{odd:} \hspace{1em} & (e_{L\ast} S ), (e_{x'} S ), (e_{L\ast}Te_{y'}), (e_{x'}Te_{y'}).
\end{align}
\end{subequations}
These properties will be used in the decomposition of the polarized cross section in
Sec.~\ref{subsec:structure_functions}.

The invariant polarization parameters have a simple interpretation in the class of
collinear frames, which contains the target rest frame as a special case (see Sec.~\ref{sec:kin}).
In the target rest frame, the polarization 4-vector and 4-tensor have only spatial components
\begin{subequations}
\label{vector_tensor_restframe}
\begin{align}
&S^0 = 0, \hspace{1em} S^i \neq 0,
\\
&T^{0\mu}, T^{\mu 0} = 0, \hspace{1em} T^{ij} \neq 0,
\end{align}
\end{subequations}
and the basis 4-vectors are given by Eqs.~(\ref{collinear_frame_hadron}) and (\ref{set2_collinear_frame}).
The contractions in Eq.~(\ref{eq:poltensor_LT}) become
\begin{subequations}
\begin{align}
(e_{L\ast} S) &= \bm{e}_z \bm{S} = S^z,
\\
-(e_{x'} S) &= \bm{e}_{x'} \bm{S} = S^{x'},
\\
-(e_{y'} S) &= \bm{e}_{y'} \bm{S} = S^{y'},
\\[.5ex]
(e_{L\ast} T e_{L\ast}) &= \bm{e}_z \bm{T} \bm{e}_z = T^{zz}, \hspace{1em} \textrm{etc.}
\end{align}
\end{subequations}
This directly connects the invariant polarization parameters with the target spin state
in the rest frame and can be used for polarization preparation (see Sec.~\ref{subsec:preparation})
and dynamical calculations (see {\secondpart}). Note that $S_L = +1$ corresponds to
spin projection $S^z = +1$ on the $z$-axis, opposite to the direction of $\bm{q}$; in agreement with the
convention used for the spin-1/2 target \cite{Bacchetta:2004jz}.
The angle $\phi_S$ is the azimuthal angle of the transverse vector polarization,
and $\phi_{T_L}$ and $\phi_{T_T}$ are the azimuthal angles characterizing
the transverse directions of the tensor polarization;
their definition conforms to the Trento convention \cite{Bacchetta:2004jz}.

In the argument of the trigonometric functions in Eq.~(\ref{eq:poltensor_LT}) the polarization angles
$\phi_S, \phi_{T_L}, \phi_{T_T}$ appear before the hadron angle $\phi_h$, as is natural when projecting
the polarization vector/tensor on the basis vectors aligned with the hadron plane. In the following we
write the arguments such that $\phi_h$ comes before the polarization angles, which is standard in
the presentation of the SIDIS cross section (see Sec.~\ref{sec:cross_sec}),
\begin{subequations}
\begin{align}
\cos(\phi_S-\phi_h) &\rightarrow \cos(\phi_h-\phi_S),
\\[1ex]
\sin(\phi_S-\phi_h) &\rightarrow -\sin(\phi_h-\phi_S), \hspace{1em} \textrm{etc.}
\end{align}
\end{subequations}
This results in the appearance of a minus sign in the structures proportional to sine functions,
which cancels the minus sign already present in Eq.~(\ref{eq:poltensor_LT}).

\subsection{Tensor representation}
The 4-tensor density matrix can be expanded in tensors formed from the
basis vectors Eqs.~(\ref{basis_longit_set2}) and (\ref{basis_transv_hadron}),
with the coefficients given by the invariant polarization parameters.
This representation is useful because it expresses the 4-dimensional density matrix
explicitly in terms of the invariant polarization parameters. It can also be connected
with the 3-dimensional cartesian components of the density matrix in the
rest frame and its angular momentum decomposition.

The 4-vector parameter $S$ in Eq.~(\ref{density_tensor_parametrization})
is expanded in the orthonormal basis vectors as
\begin{align}
S^\alpha &= -S_L e_{L\ast}^\alpha
\nonumber \\[1ex]
&+ S_T [\cos (\phi_h - \phi_S) e_{x'}^\alpha
+ \sin (\phi_h - \phi_S) e_{y'}^\alpha ].
\label{vector_expansion}
\end{align}
To expand the 4-tensor parameter $T$, we form the basis tensors
\begin{subequations}
\begin{align}
e_{LL}^{\alpha\beta} &= \frac{1}{\sqrt{6}}
(2 e_{L\ast}^\alpha e_{L\ast}^\beta - e_{x'}^\alpha e_{x'}^\beta - e_{y'}^\alpha e_{y'}^\beta),
\\
e_{LT}^{\alpha\beta}
&= \frac{1}{\sqrt{2}} ( -e_{L\ast}^\alpha e_{x'}^\beta - e_{x'}^\alpha e_{L\ast}^\beta ),
\\
e_{LT'}^{\alpha\beta}
&= \frac{1}{\sqrt{2}} ( -e_{L\ast}^\alpha e_{y'}^\beta + e_{y'}^\alpha e_{L\ast}^\beta),
\\
e_{TT}^{\alpha\beta}
&= \frac{1}{\sqrt{2}} (e_{x'}^\alpha e_{x'}^\beta - e_{y'}^\alpha e_{y'}^\beta ),
\\
e_{TT'}^{\alpha\beta}
&= \frac{1}{\sqrt{2}} (e_{x'}^\alpha e_{y'}^\beta + e_{y'}^\alpha e_{x'}^\beta ),
\end{align}
\end{subequations}
which are real, traceless, and normalized such that
\begin{align}
e^{\alpha\beta}_{I} e^{\alpha\beta}_{I'} = \delta_{II'}
\hspace{1em}
(I = LL, LT, LT', TT, TT').
\label{basis_tensor_orthogonality}
\end{align}
The tensor $T$ is expanded in these basis tensors as
\begin{align}
T^{\alpha\beta} &= \sqrt{\frac{3}{2}} \; T_{LL} e_{LL}^{\alpha\beta}
\nonumber \\
&+ \sqrt{2} \; T_{LT} \cos(\phi_h-\phi_{T_L}) e_{LT}^{\alpha\beta}
\nonumber \\[1ex]
&+ \sqrt{2} \; T_{LT} \sin(\phi_h-\phi_{T_L}) e_{LT'}^{\alpha\beta}
\nonumber \\
&+ \frac{1}{\sqrt{2}} \; T_{TT} \cos(2\phi_h-2\phi_{T_T}) e_{TT}^{\alpha\beta}
\nonumber \\
&+ \frac{1}{\sqrt{2}} \; T_{TT} \sin(2\phi_h-2\phi_{T_T}) e_{TT'}^{\alpha\beta} .
\label{tensor_expansion}
\end{align}
The expansion coefficients can be determined by contracting both sides of Eq.~(\ref{tensor_expansion})
with the basis tensors, using the orthogonality condition Eq.~(\ref{basis_tensor_orthogonality})
and the expressions of the contractions in Eq.~(\ref{eq:poltensor_LT}).

In the target rest frame the polarization 4-vector and 4-tensor have
only spatial components, Eq.~(\ref{vector_tensor_restframe}), and the
basis 4-vectors are given by Eqs.~(\ref{collinear_frame_hadron}) and (\ref{set2_collinear_frame}).
The expansion of the polarization vector, Eq.~(\ref{vector_expansion}), becomes
\begin{align}
\bm{S} &= S_L \bm{e}_{z}
\nonumber \\
&+ S_T [\cos (\phi_h - \phi_S) \bm{e}_{x'} + \sin (\phi_h - \phi_S) \bm{e}_{y'}].
\label{vector_restframe}
\end{align}
The expansion of the polarization tensor becomes (here $i,j =$ 1,2,3)
\begin{align}
T^{ij} &= \sqrt{\frac{3}{2}} \; T_{LL} e_{LL}^{ij}
\nonumber \\
&+ \sqrt{2} \; T_{LT} \cos(\phi_h-\phi_{T_L}) e_{LT}^{ij}
\nonumber \\[1ex]
&+ \sqrt{2} \; T_{LT} \sin(\phi_h-\phi_{T_L}) e_{LT'}^{ij}
\nonumber \\
&+ \frac{1}{\sqrt{2}} \; T_{TT} \cos(2\phi_h-2\phi_{T_T}) e_{TT}^{ij}
\nonumber \\
&+ \frac{1}{\sqrt{2}} \; T_{TT} \sin(2\phi_h-2\phi_{T_T}) e_{TT'}^{ij} ,
\label{tensor_restframe}
\end{align}
where the spatial components of the basis tensors are given by
\begin{subequations}
\label{basis_tensors}
\begin{align}
e_{LL}^{ij} &= \frac{1}{\sqrt{6}} (3 e_{z}^i e_{z}^j - \delta^{ij})
= \mathcal{Y}_{20}^{ij}
\\
e_{LT}^{ij}
&= \frac{1}{\sqrt{2}} ( e_{z}^i e_{x'}^j + e_{x'}^i e_{z}^j )
\nonumber \\
&= -\frac{1}{\sqrt{2}} ( \mathcal{Y}_{21}^{ij} - \mathcal{Y}_{2-1}^{ij}),
\\
e_{LT'}^{ij}
&= \frac{1}{\sqrt{2}} ( e_{z}^i e_{y'}^j + e_{y'}^i e_{z}^j)
\nonumber \\
&= \frac{i}{\sqrt{2}} ( \mathcal{Y}_{21}^{ij} + \mathcal{Y}_{2-1}^{ij}),
\\
e_{TT}^{ij}
&= \frac{1}{\sqrt{2}} (e_{x'}^i e_{x'}^j - e_{y'}^i e_{y'}^j ),
\nonumber \\
&= \frac{1}{\sqrt{2}} ( \mathcal{Y}_{22}^{ij} + \mathcal{Y}_{2-2}^{ij}),
\\
e_{TT'}^{ij}
&= \frac{1}{\sqrt{2}} (e_{x'}^i e_{y'}^j + e_{y'}^i e_{x'}^j )
\nonumber \\
&= -\frac{i}{\sqrt{2}} ( \mathcal{Y}_{22}^{ij} - \mathcal{Y}_{2-2}^{ij}).
\end{align}
\end{subequations}
Here $\mathcal{Y}_{LM}$ are the 3-dimensional spherical tensors with angular momentum $L = 2$ and
projection $M = 0, \pm 1, \pm 2$ on the $z$-axis (see e.g.\ Ref.~\cite{GonzalezLedesma:2020dgx})
\begin{subequations}
\begin{align}
\mathcal{Y}_{20} =& \frac{1}{\sqrt{6}} (3 \bm{e}_{z} \otimes \bm{e}_{z} - 1),
\\
\mathcal{Y}_{2\pm 1} =& \mp \frac{1}{\sqrt{2}}
(\bm{e}_{z} \otimes \bm{e}_{\pm} + \bm{e}_{\pm} \otimes \bm{e}_{z}),
\\[.5ex]
\mathcal{Y}_{2\pm 2} =& \bm{e}_{\pm} \otimes \bm{e}_{\pm}
\\[.5ex]
& \left[ \bm{e}_{\pm} \equiv \frac{1}{\sqrt{2}} (\bm{e}_{x'} \pm i \bm{e}_{y'}) \right],
\nonumber
\end{align}
\end{subequations}
whose projections on a unit vector $\bm{n}$ are the spherical harmonics with $L = 2$
\begin{align}
\bm{n} \cdot \mathcal{Y}_{2M} \cdot \bm{n}
= \sqrt{\frac{2 \cdot 4\pi}{5\cdot 3}} \; Y_{2M} (\bm{n}).
\label{spherical_harmonics}
\end{align}
The 4-dimensional expansion Eq.~(\ref{tensor_expansion}) thus follows the angular momentum decomposition
of the 3-dimensional polarization tensor in the target rest frame and connects it with the invariant
polarization parameters. In dynamical calculations Eq.~(\ref{tensor_expansion}) can be used to compute
the cross section proportional to a given invariant polarization parameter
(see {\secondpart}).

\subsection{Spin quantum number representation}
\label{sec:density_lambda}
From the basis vectors $e_{L\ast}$ and $e_{x'}, e_{y'}$ one can also construct a set of 4-vector
spin wave functions $\epsilon (\lambda)$ for the spin-1 particle,
\begin{subequations}
\label{wave_functions_explicit}
\begin{align}
\epsilon^\alpha (\lambda = 0) &= -e_{L\ast}^\alpha,
\\
\epsilon^\alpha (\lambda = \pm 1) &= \mp \frac{1}{\sqrt{2}} (e_{x'}^\alpha \pm i e_{y'}^\alpha).
\end{align}
\end{subequations}
In the collinear frame these vectors are aligned with the $z$ and $x', y'$ directions, and the
polarization states correspond to spin projection $0$ and $\pm 1$ on the $z$-axis (see Fig.~\ref{fig:kin}).
These vectors can be regarded as the result of a boost of the corresponding rest frame vectors,
as indicated in Eq.~(\ref{boost}).
[For collinear boosts (along the $z$-axis) canonical and light-front boosts coincide,
so there is no ambiguity in the definition of the spin wave function of a target with zero
transverse momentum; this fact will be used in the dynamical calculations using
light-front quantization in {\secondpart}.]
Using Eq.~(\ref{projection}) and the spin wave functions Eq.~(\ref{wave_functions_explicit}),
we can now compute the density matrix in the spin quantum number representation.
We obtain
\begin{widetext}
\renewcommand{\arraystretch}{1.5}
\begin{align}
 \label{eq:densitymatrix1}
\rho(\lambda,\lambda')=
\left [
\begin{array}{ccc}
\frac{1}{3}+\frac{1}{2}S_L+
\frac{1}{2}
T_{
LL}\qquad
&
\frac{1}{2\sqrt{2}}S_T
e^{-i(\phi_h-\phi_S)}\qquad
&
\frac{1}{2}
T_{TT}e^{
-i(2\phi_h-2\phi_ { T_T } )}
\\
&{}+
\frac{1}{\sqrt{2}}
T_{LT}e^{
-i(\phi_h- \phi_{T_L } ) }\qquad&
\\[10pt]
\frac{1}{2\sqrt{2}}S_T 
e^{i(\phi_h-\phi_S)}
&
\frac{1}{3}- T_{LL}
&
\frac{1}{2\sqrt{2}}S_T 
e^{-i(\phi_h-\phi_S)}
\\
{}+
\frac{1}{\sqrt{2}}
T_{LT}e^{
i(\phi_h-\phi_{T_L } ) }&&{}-
\frac{1}{\sqrt{2}}
T_{LT}e^{
-i(\phi_h- \phi_{T_L } ) }\\[10pt]
\frac{1}{2}
T_{TT}e^{
i(2\phi_h-2\phi_ { T_T } )}
&
\frac{1}{2\sqrt{2}}S_T 
e^{i(\phi_h-\phi_S)}\qquad
&
\frac{1}{3}-\frac{1}{2}S_L+
\frac{1}{2}
T_{
LL}\\
&{}-
\frac{1}{\sqrt{2}}
T_{LT}e^{
i(\phi_h - \phi_{T_L } ) }\qquad&
\end{array}
\right ].
\end{align}
\end{widetext}
Equation~(\ref{eq:densitymatrix1}) expresses the 3-dimensional density matrix in the spin quantum
numbers in terms of the intensity and angle parameters characterizing the vector and tensor polarization.
It can be used to evaluate spin observables in the spin quantum number representation
Eq.~(\ref{observable_average}).

In the treatment here we have introduced the density matrix in the 4-dimensional representation,
specified its form using its 4-dimensional properties, and then derived the 3-dimensional representation.
In Appendix~\ref{app:density_matrix} we present an alternative approach, where we
start with the specific form of the 3-dimensional density matrix and derive its 4-dimensional extension.
That formulation is useful for connecting with the treatment of spin-1 systems in nonrelativistic physics.
\section{Cross section}
\label{sec:cross_sec}
\subsection{Semi-inclusive hadronic tensor}
The differential cross section for SIDIS on a spin-1 target can be computed using
standard methods \cite{Berestetskii:1982qgu}. The hadronic structure is abstracted
in the hadronic tensor of the process, whose structural decomposition can be determined
from general considerations. We now perform the decomposition of the hadronic tensor
and derive the form of the semi-inclusive cross section for the spin-1 target.

The differential cross section for semi-inclusive scattering Eq.~(\ref{process}),
on a spin-1 target characterized by a spin density matrix $\rho$, is given by
\begin{align}
d\sigma &= \frac{1}{4I} \sum_{\lambda_{e'}\lambda_h} 
\sum_{\lambda'\lambda} \rho(\lambda, \lambda')
\nonumber \\
& \times
\sum_X \mathcal{M}(\lambda_{e'}, 
\lambda_h; \lambda_e, \lambda')^*\mathcal{M}(\lambda_{e'},\lambda_h; \lambda_e, 
\lambda) 
\nonumber \\
& \times (2\pi)^4\delta^ { (4) } (P+q-P_h-P_X)d\Gamma_ {l' }d\Gamma_{P_h} .
\label{dsigma}
\end{align}
Here
\begin{align}
I \equiv \sqrt{(P\cdot l)^2} = \tfrac{1}{2} (s - M^2)
\end{align}
is the invariant flux factor (we neglect the lepton mass).
$\mathcal{M}$ is the scattering amplitude of the process $e + A \rightarrow e' + X + h$,
where $h$ is the identified hadron and $X$ is a multi-hadron remnant state;
the amplitude is normalized according to the relativistic convention \cite{Berestetskii:1982qgu}.
In Eq.~(\ref{dsigma}) we indicate only the dependence of the amplitude on the spin variables:
the initial and final lepton helicities, $\lambda_e$ and $\lambda_{e'}$; the target spin
projection $\lambda$, and the identified hadron spin, $\lambda_h$; we do not indicate
the dependence on the particle momenta or any of the variables of the multi-hadron state.
The spin variables of the final state, $\lambda_{e'}$ and $\lambda_h$, are summed over.
The target spin projections in the initial state, $\lambda$ and $\lambda'$ in the amplitude
and its complex conjugate, are averaged over with the density matrix $\rho(\lambda,\lambda')$.
$d\Gamma_{l'}$ and $d\Gamma_{P_h}$ are the Lorentz-invariant phase-space elements of the
final lepton and the identified hadron,
\begin{subequations}
\begin{align}
&d\Gamma_{l'} \equiv \frac{d^4 l'}{(2\pi)^4} \; \delta(l^{\prime 2}) \theta (l^{\prime 0}) 
\; = \;
\frac{d^3l'}{2|\bm{l}'| (2\pi)^3},
\label{dGamma_electron}
\\
&d\Gamma_{P_h} \equiv \frac{d^4 P_h}{(2\pi)^4} \; \delta(P_h^2 - M_h^2) \theta (P_h^0) 
\; = \;
\frac{d^3P_h}{2E_h (2\pi)^3}.
\label{eq:dG_h}
\end{align}
\end{subequations}
The summation over $X$ in Eq.~(\ref{dsigma}) denotes both the integration over the variables
of the individual multi-hadron remnant state $X$ and the summation over all the
allowed states. The $\delta$ function ensures 4-momentum (energy and momentum)
conservation.\footnote{In the simplified derivation here we assume that the scattering events
contributing to the semi-inclusive cross section have exactly one hadron $h$ in the
phase space element at the given momentum $P_h$, and that the hadrons in the remnant state
$X$ are kinematically distinct from the identified hadron. This allows us to express the
cross section in terms of the scattering amplitudes without accounting for the hadron
multiplicity. The results for the cross section decomposition are not affected by this simplification;
the representation in terms of amplitudes is used only to explain the origin of the hadronic tensor.
The amplitude-level representation of the cross section can be generalized to account for
final states with multiple hadrons $h$; see e.g.\ Ref.~\cite{Donnachie:1978oje}.}

In leading order of the electromagnetic coupling (one-photon exchange approximation)
the scattering amplitude is given by the product of the electromagnetic currents of the 
lepton and hadron transitions,
\begin{align}
& \mathcal{M}(\lambda_{e'},\lambda_h; \lambda_e, \lambda)
\nonumber \\
&= -\frac{e^2}{Q^2} \
\langle e'(\lambda_{e'}) | J_\mu | e(\lambda_e) \rangle \;
\langle X, h(\lambda_{h}) | J^\mu | A(\lambda) \rangle ,
\end{align}
where $e$ is the elementary charge. The cross section can then be expressed in terms
of a leptonic and hadronic scattering tensor in the standard fashion.
The  leptonic tensor is defined as
\begin{subequations}
\begin{align}
&L^{\mu\nu} \equiv \sum_{\lambda_{e'}} \langle e(\lambda_{e}) | J^{\dagger\mu} | 
e'(\lambda_{e'}) \rangle 
\langle e'(\lambda_{e'}) | J^\nu | e(\lambda_e) \rangle 
\\[1ex]
&= 4l^\mu l^\nu-Q^2g^{\mu\nu}-2l^{\{\mu}q^{\nu\}}+(2\lambda_e)2i\epsilon^{
\mu\nu\rho\sigma}q_\rho l_\sigma ,
\label{leptonic_tensor}
\end{align}
\end{subequations}
where $\lambda_{e'} = \lambda_e$ in the sum (the lepton helicity
is conserved by the interaction because we neglect the lepton mass).
Here and in the following we use
$\{...\}$ and $[...]$ to denote symmetrization and antisymmetrization in Lorentz 
indices,
\begin{equation}
a^{\{\mu}b^{\nu\}} = a^\mu b^\nu + b^\nu a^\mu,
\hspace{1em}
a^{[\mu}b^{\nu]} = a^\mu b^\nu - b^\nu a^\mu .
\end{equation}
One refers to the symmetric, lepton helicity-independent and antisymmetric,
lepton-helicity-dependent parts as the unpolarized and polarized parts
of the leptonic tensor. The hadronic tensor is defined as
\begin{align}
& W^{\mu\nu}(\lambda',\lambda)
\equiv \frac{1}{4\pi} \sum_{\lambda_h} \sum_X \; (2\pi)^4\delta^4(P+q-P_h-P_X)
\nonumber\\
& \times \langle A(\lambda')|J^{\dagger \mu}|X, h(\lambda_h)\rangle \langle X, 
h(\lambda_h)| J^\nu|A(\lambda) \rangle .
\label{eq:hadronictensor}
\end{align}
Notice that the tensor is a matrix in the spin quantum numbers of the target states in the
current matrix element and its complex conjugate ($\lambda' \neq \lambda$).
It is assumed that the target spin states are prepared in the same way as those
in the density matrix (see Sec.~\ref{sec:densitymatrix}). With these definitions one has
\begin{align} 
&\sum_{\lambda_{e'}\lambda_h} 
\sum_{\lambda'\lambda} \rho(\lambda, \lambda')
\nonumber \\
&\times \sum_X \mathcal{M}(\lambda_{e'}, 
\lambda_h; \lambda_e, \lambda')^*\mathcal{M}(\lambda_{e'},\lambda_h; \lambda_e, 
\lambda) 
\nonumber\\
&\times (2\pi)^4\delta^ { (4) } (P+q-P_h-P_X)d\Gamma_ {l' }d\Gamma_{P_h}
\nonumber \\
&= \frac{4\pi e^4}{Q^4} \; L_{\mu\nu} \langle W^{\mu\nu} \rangle \,,
\end{align}
where
\begin{equation}
\langle W^{\mu\nu} \rangle \equiv 
\sum_{\lambda'\lambda}\rho(\lambda,\lambda')W^{\mu\nu}(\lambda',\lambda)
\label{eq:hadronictensor_contracted}
\end{equation}
is the hadronic tensor of the polarized target, averaged over the target spin states with the
spin density matrix, see Eq.~(\ref{observable_average}).

The phase space element of the final lepton, Eq.~(\ref{dGamma_electron}), can be expressed
as a differential in the invariants $x_A$ and $Q^2$, and in the azimuthal angle of the final
lepton around the initial lepton momentum direction in the target rest frame, $\psi_{l'}$,
defined with regard to a reference direction perpendicular to the initial lepton momentum;
the calculations are identical to those in unpolarized SIDIS and need
not be repeated here \cite{Bacchetta:2006tn}. Altogether the differential cross section
is obtained as
\begin{equation} 
d\sigma = \frac{2\pi y^2\alpha_{\rm em}^2}{Q^6} \; dx_A dQ^2 \frac{d\psi_{l'}}{2\pi} \;
L_{\mu\nu}\langle W^{\mu\nu} \rangle \; d\Gamma_{P_h} ,
\label{eq:crossinterm}
\end{equation}
where $\alpha_{\rm em} = e^2/(4\pi) \approx 1/137$ is the electromagnetic fine 
structure constant. The form of Eq.~(\ref{eq:crossinterm}) separates the kinematic variables
determined by the final lepton (which are the same as in inclusive scattering) from the
variables characterizing the hadronic final state.

\subsection{Semi-inclusive structure functions}
\label{subsec:structure_functions}

The semi-inclusive hadronic tensor Eq.~(\ref{eq:hadronictensor_contracted})
has a rich structure, in which the target polarization in the initial state is entangled
with the hadron momentum in the final state. It can be expanded in kinematic tensors
formed from the 4-vectors of the scattering process (see Sec.~\ref{subsec:basis_vectors}),
multiplied by the invariant polarization parameters characterizing the target polarization
(see Sec.~\ref{subsec:invariant_polarization}).
The contraction of these kinematic tensors with the leptonic tensor then determines
the $y$ and $\phi_h$ dependence of the cross section. The invariant polarization parameters
accompanying the structures determine the target spin dependence. The invariant coefficients
in this expansion are the polarized semi-inclusive structure functions, which describe
the dynamics of the scattering process. We now perform this expansion, determine the
independent structures, and exhibit the kinematic dependences of the cross section.

In the first step, we enumerate the possible kinematic tensors that can appear in the
expansion of the hadronic tensor. The kinematic tensors must obey the transversality
conditions on the hadronic tensor,
\begin{equation}
q_\mu \langle W^{\mu\nu} \rangle = 0, \hspace{2em} \langle W^{\mu\nu} \rangle q_\nu = 0\,.
\label{eq:transversality}
\end{equation}  
In addition, they must satisfy the hermiticity conditions
\begin{equation}
\langle W^{\{\mu\nu\}} \rangle \;\; \textrm{real}, \hspace{2em} \langle W^{[\mu\nu]} \rangle
\;\; \textrm{imaginary} ,
\label{hermiticity}
\end{equation}  
which follow from the fact that the symmetric (antisymmetric) part of the leptonic tensor
Eq.~(\ref{leptonic_tensor}) is real (imaginary)
and must produce a real number upon contraction with the hadronic tensor. It is natural to build the
kinematic tensors from the longitudinal basis vectors of Set I, Eq.~(\ref{basis_longit_set1}),
for which the transversality conditions Eq.~(\ref{eq:transversality}) take a simple form,
and from the transverse basis vectors aligned with the hadron momentum, Eq.~(\ref{basis_transv_hadron}).
We obtain nine kinematic tensors of different symmetry and parity:
\begin{itemize}
\item four symmetric tensors:
\begin{align}
&\mathcal{A}_L^{\mu\nu}
=\tfrac{1}{2}e_L^\mu e_L^\nu\,,
&\mathcal{A}_T^{\mu\nu}
=\tfrac{1}{2} ( e_{x'}^{\mu} e_{x'}^{\nu} + e_{y'}^{\mu} e_{y'}^{\nu}) ,
\nonumber\\
&\mathcal{A}_{LT}^{\mu\nu}
=-\tfrac{1}{2} e_L^{\{\mu} e_{x'}^{\nu\}},
&\mathcal{A}_{TT}^{\mu\nu}
=\tfrac{1}{2} (e_{x'}^{\mu} e_{x'}^{\nu} - e_{y'}^{\mu} e_{y'}^{\nu});
\label{eq:symm_tensor}
\end{align}
\item one antisymmetric tensor:
\begin{equation} 
\mathcal{A}_{LT'}^{\mu\nu}
= \tfrac{i}{2}\epsilon^{\mu\nu\rho\sigma} e_{q, \rho}  e_{y',\sigma}
= \tfrac{i}{2} e_L^{[\mu} e_{x'}^{\nu]};
\label{eq:asymm_tensor}
\end{equation}
\item
two symmetric pseudotensors:
\begin{align}
&\mathcal{A}_{LT''}^{\mu\nu}
= -\tfrac{1}{2} e_L^{\{\mu} e_{y'}^{\nu\}},
&\mathcal{A}_{TT'}^{\mu\nu}
= +\tfrac{1}{2} e_{x'}^{\{\mu} e_{y'}^{\nu\}};
\label{eq:symm_pstensor}
\end{align}
\item two antisymmetric pseudotensors:
\begin{align} 
&\mathcal{A}_{LT'''}^{\mu\nu}
= \tfrac{i}{2} \epsilon^{\mu\nu\rho\sigma} e_{q,\rho} e_{x',\sigma}
= -\tfrac{i}{2} e_L^{[\mu} e_{y'}^{\nu]}\,,
\nonumber \\
&\mathcal{A}_{TT''}^{\mu\nu}
=-\tfrac{i}{2} \epsilon^{\mu\nu\rho\sigma} e_{q,\rho} e_{L,\sigma}
= \tfrac{i}{2} e_{x'}^{[\mu} e_{y'}^{\nu]}.
\label{eq:asymm_pstensor}
\end{align}
\end{itemize}
The symmetric tensors will combine with the unpolarized part of the 
leptonic tensor; the antisymmetric tensors will combine with the polarized part.

In the next step, we multiply the kinematic tensors with the invariant polarization parameters
representing the possible dependencies on the target polarization arising from the average
with the density matrix. The available parameters are unity (no polarization) and the three vector-
and five tensor-polarized parameters of Eq.~(\ref{eq:poltensor_LT}), whose parity is given
in Eq.~(\ref{polarizations_parity}). Altogether, we have as possible parameters five scalars:
\begin{itemize}
 \item one unpolarized: 1 (\text{unity});
 \item one vector-polarized: $S_T\sin(\phi_h-\phi_S)$;
 \item three tensor-polarized: $T_{LL}$, $ T_{LT}\cos(\phi_h-\phi_{T_L})$, \\
 $T_{TT}\cos(2\phi_h-2\phi_{T_T})$;
\end{itemize}
and four pseudoscalars:
\begin{itemize}
 \item two vector-polarized: $S_L, S_T
\cos(\phi_h-\phi_S)$;
\item two tensor-polarized:
$T_{LT}\sin(\phi_h-\phi_{T_L})$; \\
$T_{TT}\sin(2\phi_h-2\phi_{T_T})$.
\end{itemize}
When multiplying these parameters with the kinematic tensors we must take into account
the condition of overall parity
\begin{align}
\langle W^{\mu\nu} \rangle = \; \textrm{true tensor (parity-even).}
\end{align}
So the parity-even kinematic tensors in Eqs.~(\ref{eq:symm_tensor}) and (\ref{eq:asymm_tensor})
can only be multiplied with the scalar parameters, while the parity-odd kinematic tensors
in Eqs.~(\ref{eq:symm_pstensor})-(\ref{eq:asymm_pstensor}) can only be multiplied with
the pseudoscalar parameters. Altogether this results in
\begin{align*}
5\times5+4\times4=41
\end{align*}
independent structures in the polarized hadronic tensor.

Each independent structure is multiplied by a structure function describing the
dynamical content. We use the following notation:
\begin{align}
&F^{[\text{azimuthal}]}_{[\text{beam}][\text{target}],[\text{photon}]}
\nonumber \\[2ex]
\text{beam}&=U,L\nonumber\\
\text{target}&=U,S_L,S_T,T_{LL},T_{LT},T_{TT}\nonumber\\
\text{azimuthal}&= 1, \cos (n\phi_h\pm\phi_i), \sin 
(n\phi_h\pm\phi_i)\nonumber\\
\text{photon}&=L,T
\end{align}
The labels anticipate properties of the terms in the cross section that will appear
after contraction with the leptonic tensor (see below). The label \emph{beam} refers to the
dependence on the lepton helicity (symmetric or antisymmetric part of the hadronic tensor);
\emph{target} refers to the dependence on the target polarization (polarization parameters in hadronic tensor);
\emph{photon} denotes the polarization of the virtual photon (longitudinal or transverse part
of the hadronic tensor); \emph{azimuthal} refers to the dependence on the hadron azimuthal
angle $\phi_h$ that appears after contraction with the leptonic tensor.
The structure functions depend on four kinematic variables: the invariant DIS variables $x_A$ and $Q^2$,
Eq.~(\ref{scaling_variables_general}); the invariant modulus of the final-state hadron transverse momentum
$|\bm P_{hT}|$, Eq.~(\ref{P_hT_invariant}); and one variable characterizing the longitudinal momentum,
e.g.\ $z_h$ or $\zeta_h$, Eqs.~(\ref{z_h_def}) or (\ref{zeta_h_def}) (see Sec.~\ref{subsec:final_state}),
\begin{align}
F \equiv F(x_A, Q^2; z_h, |\bm P_{hT}|).
\end{align}
The dependence of the cross section on $\phi_h$ is explicit and follows from the decomposition of
the hadronic tensor (see below).

We can now enumerate the 41 possible structures in the polarized hadronic tensor
Eq.~(\ref{eq:hadronictensor_contracted}), including their structure functions.
They are:

\begin{widetext}
\begin{itemize}
 \item four with lepton unpolarized, target unpolarized:
 \begin{align}
F_{UU,L}\mathcal{A}^{\mu\nu}_L, \quad
F_{UU,T}\mathcal{A}^{\mu\nu}_T, \quad
F_{UU}^{\cos\phi_h}\mathcal{A}^{\mu\nu}_{LT},\quad
F_{UU}^{\cos2\phi_h}\mathcal{A}^{\mu\nu}_{TT};
\label{eq:41sf_first}
\end{align}

\item one with lepton polarized, target unpolarized:
\begin{equation}
F_{LU}^{\sin\phi_h}\mathcal{A}^{\mu\nu}_{LT'};
\end{equation}

\item two with lepton unpolarized, target vector-polarized longitudinally:
\begin{equation}
S_L F_{US_L}^{\sin\phi_h}\mathcal{A}^{\mu\nu}_{LT''}
,\quad 
S_L F_{US_L}^{\sin2\phi_h} \mathcal{A}^{\mu\nu}_{TT'};
\end{equation}

\item two with lepton polarized, target vector-polarized longitudinally:
\begin{equation}
S_L F_{LS_L}\mathcal{A}^{\mu\nu}_{ TT''},\quad
S_L F_{LS_L}^{\cos\phi_h} 
\mathcal{A}^{\mu\nu}_{LT'''};
\end{equation}

\item six with lepton unpolarized, target vector-polarized transversely:
\begin{align}
&S_T\sin(\phi_h-\phi_S)F_{US_T,L}^{\sin(\phi_h-\phi_S)}\mathcal{A}^
{\mu\nu}_{L},
\nonumber \\
&S_T\sin(\phi_h-\phi_S)F_{US_T,T}^{\sin(\phi_h-\phi_S)} 
\mathcal{A }^{\mu\nu}_{T},\nonumber\\ 
&S_T F_{US_T}^{\sin(\phi_h+\phi_S)}\left[
\cos(\phi_h-\phi_S)\mathcal{A}^{\mu\nu}_{TT'}-
\sin(\phi_h-\phi_S)\mathcal{A}^{\mu\nu}_{TT}\right],\nonumber\\
&S_T F_{US_T}^{\sin(3\phi_h-\phi_S)}\left[
\cos(\phi_h-\phi_S)\mathcal{A}^{\mu\nu}_{TT'}+
\sin(\phi_h-\phi_S)\mathcal{A}^{\mu\nu}_{TT}\right],\nonumber\\
&S_T F_{US_T}^{\sin\phi_S} \left[
\cos(\phi_h-\phi_S)\mathcal{A}^{\mu\nu}_{LT''}- 
\sin(\phi_h-\phi_S)\mathcal{A}^{\mu\nu}_{LT}\right],\nonumber\\
&S_T F_{US_T}^{\sin(2\phi_h-\phi_S)} \left[
\cos(\phi_h-\phi_S)\mathcal{A}^{\mu\nu}_{LT''}+ 
\sin(\phi_h-\phi_S)\mathcal{A}^{\mu\nu}_{LT}\right];
\end{align}

\item three with lepton polarized, target vector-polarized transversely:
\begin{align}
&S_T\cos(\phi_h-\phi_S)F_{LS_T}^{\cos(\phi_h-\phi_S)}
\mathcal{A}^{\mu\nu}_{TT''},\nonumber\\
&S_T F_{LS_T}^{\cos\phi_S}
\left[ 
\cos(\phi_h-\phi_S)\mathcal{A}^{\mu\nu}_{ LT'''}+
\sin(\phi_h-\phi_S)\mathcal{A}^{\mu\nu}_{LT'}
\right],
\nonumber\\
&S_T F_{LS_T}^{\cos(2\phi_h-\phi_S)}\left[
\cos(\phi_h-\phi_S)\mathcal{A}^{\mu\nu}_{ LT'''}-
\sin(\phi_h-\phi_S)\mathcal{A}^{\mu\nu}_{LT'}
\right];
\end{align}

\item four with lepton unpolarized, target tensor-polarized longitudinally:
\begin{equation}
 T_{LL}F_{U T_{LL},L}
\mathcal{A}^{\mu\nu}_{L},\quad 
 T_{LL}F_{U T_{LL},T}
\mathcal{A}^{\mu\nu}_{T},\quad
 T_{LL}F_{U T_{LL}}^{\cos\phi_h}
\mathcal{A}^{\mu\nu}_{LT},\quad
 T_{LL}F_{U T_{LL}}^{\cos2\phi_h}
\mathcal{A}^{\mu\nu}_{TT};
\end{equation}

\item one with lepton polarized, target tensor-polarized longitudinally:
\begin{equation}
 T_{LL}F_{L T_{LL}}^{\sin\phi_h}
\mathcal{A}^{\mu\nu}_{LT'};
\end{equation}

\item six with lepton unpolarized, target tensor-polarized longitudinal-transversely:
\begin{align}
& T_{LT}\cos(\phi_h-\phi_{T_L})
F_{U T_{LT},L}^{\cos(\phi_h-\phi_{T_L})} 
\mathcal{A}^{\mu\nu}_{L},
\nonumber \\
& T_{LT}\cos(\phi_h-\phi_{T_L})
F_{U T_{LT},T}^{\cos(\phi_h-\phi_{T_L})}
\mathcal{A}^{\mu\nu}_{T}, \nonumber\\
& T_{LT} 
F_{U T_{LT}}^{\cos(\phi_h+\phi_{T_L})}\left[
\cos(\phi_h-\phi_{T_L})\mathcal{A}^{\mu\nu}_{TT}+
\sin(\phi_h-\phi_{T_L})\mathcal{A}^{\mu\nu}_{TT'}\right],\nonumber\\
& T_{LT} 
F_{U T_{LT}}^{\cos(3\phi_h-\phi_{T_L})}\left[
\cos(\phi_h-\phi_{T_L})\mathcal{A}^{\mu\nu}_{TT}-
\sin(\phi_h-\phi_{T_L})\mathcal{A}^{\mu\nu}_{TT'}\right],\nonumber\\
& T_{LT} F_{U T_{LT}}^{\cos\phi_{T_L}} 
\left[
\cos(\phi_h-\phi_{T_L})\mathcal{A}^{\mu\nu}_{LT}+
\sin(\phi_h-\phi_{T_L})\mathcal{A}^{\mu\nu}_{LT''}\right],\nonumber\\
& T_{LT} 
F_{U T_{LT}}^{\cos(2\phi_h-\phi_{T_L})} \left[
\cos(\phi_h-\phi_{T_L})\mathcal{A}^{\mu\nu}_{LT}- 
\sin(\phi_h-\phi_{T_L})\mathcal{A}^{\mu\nu}_{LT''}\right];
\end{align}

\item three with lepton polarized, target tensor-polarized longitudinal-transversely:
\begin{align}
& T_{LT}\sin(\phi_h-\phi_{T_L})F_{L T_{LT}
}^{
\sin(\phi_h-\phi_{T_L})}
\mathcal{A}^{\mu\nu}_{TT''},\nonumber\\
& T_{LT}F_{L T_{LT}}^{\sin\phi_{T_L}}
\left[ 
\cos(\phi_h-\phi_{T_L})\mathcal{A}^{\mu\nu}_{ LT'}-
\sin(\phi_h-\phi_{T_L})\mathcal{A}^{\mu\nu}_{LT'''}
\right],
\nonumber\\
& T_{LT}F_{L T_{LT}}^{\sin(2\phi_h-\phi_{
T_L})}\left[
\cos(\phi_h-\phi_{T_L})\mathcal{A}^{\mu\nu}_{ LT'}+
\sin(\phi_h-\phi_{T_L})\mathcal{A}^{\mu\nu}_{LT'''}
\right];
\end{align}

\item six with lepton unpolarized, target tensor-polarized transversely:
\begin{align}
& T_{TT}\cos(2\phi_h-2\phi_{T_T})
F_{U T_{TT},L}^{\cos(2\phi_h-2\phi_{T_T})} 
\mathcal{A}^{\mu\nu}_{L},
\nonumber \\
& T_{TT}\cos(2\phi_h-2\phi_{T_T})
F_{U T_{TT},T}^{\cos(2\phi_h-2\phi_{T_T})}
\mathcal{A}^{\mu\nu}_{T}, \nonumber\\
& T_{TT} F_{U T_{TT}}^{\cos2\phi_{T_T}}\left[
\cos(2\phi_h-2\phi_{T_T})\mathcal{A}^{\mu\nu}_{TT}+
\sin(2\phi_h-2\phi_{T_T})\mathcal{A}^{\mu\nu}_{TT'}\right],\nonumber\\
& T_{TT} 
F_{U T_{TT}}^{\cos(4\phi_h-2\phi_{T_T})}\left[
\cos(2\phi_h-2\phi_{T_T})\mathcal{A}^{\mu\nu}_{TT}-
\sin(2\phi_h-2\phi_{T_T})\mathcal{A}^{\mu\nu}_{TT'}\right],\nonumber\\
& T_{TT} F_{U T_{TT}}^{\cos(\phi_h-2\phi_{T_T})} 
\left[
\cos(2\phi_h-2\phi_{T_T})\mathcal{A}^{\mu\nu}_{LT}+
\sin(2\phi_h-2\phi_{T_T})\mathcal{A}^{\mu\nu}_{LT''}\right],\nonumber\\
& T_{TT} F_{U T_{TT}}^{\cos(3\phi_h-2\phi_{T_T})} 
\left[
\cos(2\phi_h-2\phi_{T_T})\mathcal{A}^{\mu\nu}_{LT}- 
\sin(2\phi_h-2\phi_{T_T})\mathcal{A}^{\mu\nu}_{LT''}\right];
\end{align}

\item three with lepton polarized, target tensor-polarized transversely:
\begin{align}
& T_{TT}\sin(2\phi_h-2\phi_{T_T})F_{L T_{TT}}^{
\sin(2\phi_h-2\phi_{T_T})}
\mathcal{A}^{\mu\nu}_{TT''},\nonumber\\
& T_{TT} F_{L T_{TT}}^{\sin(\phi_h-2\phi_{T_T})}
\left[ 
\sin(2\phi_h-2\phi_{T_T})\mathcal{A}^{\mu\nu}_{ LT'''}-
\cos(2\phi_h-2\phi_{T_T})\mathcal{A}^{\mu\nu}_{LT'}
\right],
\nonumber\\
& T_{TT} 
F_{L T_{TT}}^{\sin(3\phi_h-2\phi_{T_T})}\left[
\sin(2\phi_h-2\phi_{T_T})\mathcal{A}^{\mu\nu}_{ LT'''}+
\cos(2\phi_h-2\phi_{T_T})\mathcal{A}^{\mu\nu}_{LT'}
\right].
\label{eq:41sf_last}
\end{align}
\end{itemize}
\end{widetext}
The first 18 structure functions (target unpolarized and vector-polarized) also appear in
SIDIS from a spin-1/2 target~\cite{Diehl:2005pc,Bacchetta:2006tn,Bastami:2018xqd,Donnelly:2023rej};
the last 23 (target tensor-polarized) are unique to the spin-1 case.

An alternative derivation of the independent structures can be performed by projecting the
hadronic tensor on photon helicity states and using the symmetry properties of helicity amplitudes;
see Appendix \ref{sec:helicity_amps}. It confirms the number of 41 independent structures and
provides explicit expressions of the structure functions in terms of helicity amplitudes.

In the last step, we contract the independent structures in the hadronic tensor with the
leptonic tensor Eq.~(\ref{leptonic_tensor}). This produces the independent scalar terms
in the cross section with a definite $\phi_h$ dependence. It is useful to introduce the
parameter describing the ratio of longitudinal to transverse polarization of the virtual photon,
\begin{align}
\epsilon
&\equiv \frac{\phantom{-} g_L^{\mu\nu} L_{\mu\nu}}{- g_T^{\mu\nu} L_{\mu\nu}}
=
\frac{e_L^\mu e_L^\nu L_{\mu\nu}}{(e_L^\mu e_L^\nu-g^{\mu\nu}) L_{\mu\nu}}
\nonumber \\[1ex]
&= \frac{e_L^\mu e_L^\nu L_{\mu\nu}}{(e_{x'}^\mu e_{x'}^\nu+ e_{y'}^\mu e_{y'}^\nu) L_{\mu\nu}},
\end{align}
which is expressed in terms of the scaling variables as
\begin{align}
\epsilon
= \frac{1 -y -\gamma^2 y^2/4}{1 - y + y^2/2 + \gamma^2 y^2/4}.
\label{epsilon_from_y}
\end{align}
The contractions with the initial lepton momentum are then evaluated
using the relations
\begin{subequations}
\begin{align}
(e_L l) &= \frac{Q}{2}\sqrt{\frac{1+\epsilon}{1-\epsilon}}\,,
\\
(e_{x'} l)^2 + (e_{y'} l)^2 = -l_T^2 &= \frac{Q^2}{2}\frac{\epsilon}{1-\epsilon} ,
\\
 (e_{x'} l) = -{\textstyle \sqrt{-l_T^2}} \cos\phi_h
&= -\frac{Q}{2} \sqrt\frac{2 \epsilon}{1 - \epsilon} \, \cos\phi_h,
\\
(e_{y'} l) = -{\textstyle \sqrt{-l_T^2}} \sin\phi_h
&= -\frac{Q}{2} \sqrt\frac{2 \epsilon}{1 - \epsilon} \, \sin\phi_h ;
\end{align}
\end{subequations}
the last two relations follow from Eqs.~(\ref{basis_transv_lepton}), (\ref{basis_transv_hadron}),
and (\ref{basis_transv_angle}). The contraction of the leptonic tensor with the
nine kinematic tensors of Eqs.~(\ref{eq:symm_tensor})--(\ref{eq:asymm_pstensor}) gives
\begin{subequations}
\label{eq:L_contraction}
\begin{alignat}{2}
& L_{\mu\nu} \mathcal{A}_L^{\mu\nu}
&&=\frac{Q^2}{1-\epsilon}\epsilon\,,
\\
& L_{\mu\nu} \mathcal{A}_T^{\mu\nu}
&&=\frac{Q^2}{1-\epsilon}\,,
\\
& L_{\mu\nu} \mathcal{A}_{TT''}^{\mu\nu}
&&=\frac{Q^2}{1-\epsilon}\lambda_e\sqrt{1-\epsilon^2}\,,
\\
& L_{\mu\nu} \mathcal{A}_{LT}^{\mu\nu}
&&=\frac{Q^2}{1-\epsilon}\sqrt{2\epsilon(1+\epsilon) }\cos\phi_h \,,
\\
& L_{\mu\nu} \mathcal{A}_{LT''}^{\mu\nu}
&&=\frac{Q^2}{1-\epsilon}\sqrt{2\epsilon(1+\epsilon)}\sin\phi_h\,,
\\
& L_{\mu\nu} \mathcal{A}_{TT}^{\mu\nu}
&&=\frac{Q^2}{1-\epsilon}\epsilon\cos2\phi_h\,,
\\
& L_{\mu\nu} \mathcal{A}_{TT'}^{\mu\nu}
&&=\frac{Q^2}{1-\epsilon}\epsilon\sin2\phi_h\,,
\\
& L_{\mu\nu} \mathcal{A}_{LT'''}^{\mu\nu}
&&=\frac{Q^2}{1-\epsilon}\lambda_e\sqrt{2\epsilon(1-\epsilon) }\cos\phi_h\,,
\\
& L_{\mu\nu} \mathcal{A}_{LT'}^{\mu\nu}
&&=\frac{Q^2}{1-\epsilon}\lambda_e\sqrt{2\epsilon(1-\epsilon) }\sin\phi_h\,.
\end{alignat}
\end{subequations}
Using these expressions we can compute the contraction of the 41 independent tensor
structures in the hadronic tensor and assemble the terms in the cross section.

\subsection{Semi-inclusive cross section}
\label{subsec:complete}
We are now ready to write down the complete differential cross section for
semi-inclusive scattering on a polarized spin-1 target, with the 41 independent structures
identified above. We write it in the form
\begin{align}
d\sigma &= \frac{2\pi y^2\alpha_{\rm em}^2}{Q^4 (1 - \epsilon)} \; dx_A dQ^2 \frac{d\psi_{l'}}{2\pi}
\nonumber \\
&\times \left( \mathcal{F}_U+\mathcal{F}_S+\mathcal{F}_T \right) \; d\Gamma_{P_h} ,
\label{cross_section_final}
\end{align}
where $\mathcal{F}_{U, S, T}$ contain the terms with a given dependence on the
target polarization (unpolarized, vector-polarized, tensor-polarized).
The terms independent of the target polarization are
\begin{align}
\mathcal{F}_U &=F_{UU,T} + \epsilon F_{UU,L}
\nonumber \\
&+\sqrt{2\epsilon(1+\epsilon)}\cos\phi_h 
F_{UU}^{\cos\phi_h}
\nonumber\\
&+\epsilon\cos2\phi_h F_{UU}^{\cos2\phi_h} 
\nonumber \\
&+(2\lambda_e)\sqrt{2\epsilon(1-\epsilon)}\sin\phi_h 
F_{LU}^{\sin\phi_h}\,.
\label{cross_section_unpol}
\end{align}
The terms dependent on the vector polarization are
\begin{align}
&\mathcal{F}_S
\nonumber \\
&= S_L \left[
\sqrt{2\epsilon(1+\epsilon)}\sin\phi_h F_{US_L}^{\sin\phi_h}
+\epsilon \sin 2\phi_h F_{US_L}^{\sin 2\phi_h} \right]
\nonumber\\ 
&+S_L (2\lambda_e) 
\left[\sqrt{1-\epsilon^2}F_{LS_L}+ \sqrt{2\epsilon(1-\epsilon)} 
\cos\phi_h F_{LS_L}^{\cos\phi_h} \right]
\nonumber\\
&+ S_T \left[ 
\sin(\phi_h-\phi_S)\left(F_{US_T,T}^{\sin(\phi_h-\phi_S)}+\epsilon  
F_{US_T,L}^{\sin(\phi_h-\phi_S)} \right) \right.
\nonumber \\
&\hspace{2em} + \epsilon \sin(\phi_h+\phi_S) F_{US_T}^{\sin(\phi_h+\phi_S)}
\nonumber\\
& \hspace{2em} + \epsilon \sin(3\phi_h-\phi_S) 
F_{US_T}^{\sin(3\phi_h-\phi_S)}
\nonumber\\
& \hspace{2em}
+\sqrt{2\epsilon(1+\epsilon)} \sin\phi_S F_{US_T}^{\sin\phi_S}
\nonumber\\
&\left. \hspace{2em}
+ \sqrt{2\epsilon(1+\epsilon)} \sin(2\phi_h-\phi_S) 
F_{US_T}^{\sin(2\phi_h-\phi_S)} \right]
\nonumber\\
&+S_T (2\lambda_e)
\left[\sqrt{1-\epsilon^2}\cos(\phi_h-\phi_S)F_{LS_T}^{
\cos(\phi_h-\phi_
S)} \right.
\nonumber\\
& \hspace{2em}
+ \sqrt{2\epsilon(1-\epsilon)}
\cos\phi_S F_{LS_T}^{\cos\phi_S} 
\nonumber\\
& \hspace{2em}
+ \left.\sqrt{2\epsilon(1-\epsilon)}
\cos(2\phi_h-\phi_S) 
F_{LS_T}^{\cos(2\phi_h-\phi_S)} \right] \,.
\label{cross_section_vector}
\end{align}
The terms dependent on the tensor polarization are
\begin{align}
& \mathcal{F}_T
\nonumber \\
&= T_{LL} \left[  F_{U T_{LL},T} + \epsilon F_{U T_{LL},L} \phantom{0^{0^0}_{0_0}}
\right.
\nonumber \\
& \hspace{2em} +\sqrt{2\epsilon(1+\epsilon)} \cos\phi_h F_{U T_{LL}}^{\cos\phi_h}
\nonumber \\
& \left. \hspace{2em} + \epsilon\cos2\phi_h 
F_{U T_{LL}}^{\cos2\phi_h}\right]
\nonumber\\
&+ T_{LL}(2\lambda_e) 
\sqrt{2\epsilon(1-\epsilon)}\sin\phi_h 
F_{L T_{LL}}^{\sin\phi_h}
\nonumber \\
&+  T_{LT} \left[ 
\cos(\phi_h-\phi_{T_L}) \phantom{0^{0^0}_{0_0}} \right.
\nonumber \\
& \hspace{2em} \times \left( F_{U T_{LT},T}^{\cos(\phi_h-\phi_{T_L})}
+\epsilon F_{U T_{LT},L}^{\cos(\phi_h-\phi_{T_L})} \right)  
\nonumber \\
& \hspace{2em}
+ \epsilon \cos(\phi_h+\phi_{T_L}) F_{U T_{LT}}^{\cos(\phi_h+\phi_{T_L})}
\nonumber \\
& \hspace{2em}
+ \epsilon \cos(3\phi_h-\phi_{T_L}) F_{U T_{LT}}^{\cos(3\phi_h-\phi_{T_L})} 
\nonumber \\
& \hspace{2em}
+ \sqrt{2\epsilon(1+\epsilon)} \cos\phi_{T_L} F_{U T_{LT}}^{\cos\phi_{T_L}}
\nonumber \\
& \hspace{2em} 
+ \left. \sqrt{2\epsilon(1+\epsilon)} \cos(2\phi_h-\phi_{T_L}) 
F_{U T_{LT}}^{\cos(2\phi_h-\phi_{T_L})} 
\right]
\nonumber\\
& + T_{LT} (2\lambda_e) 
\left[\sqrt{1-\epsilon^2}\sin(\phi_h-\phi_{T_L}) F_{L T_{LT}}^{\sin(\phi_h-\phi_{T_L})} \right.
\nonumber\\
& \hspace{2em}
+ \sqrt{2\epsilon(1-\epsilon)} \sin\phi_{T_L} F_{L T_{LT}}^{\sin\phi_{T_L}}
\nonumber\\
& \hspace{2em}
+ \left. \sqrt{2\epsilon(1-\epsilon)} \sin(2\phi_h-\phi_{T_L}) F_{L T_{LT}}^{\sin(2\phi_h-\phi_{T_L})}
\right] 
\nonumber \\
&+  T_{TT} \left[ \cos(2\phi_h-2\phi_{T_T}) \phantom{0^{0^0}_{0_0}} \right.
\nonumber \\
& \hspace{2em} \times \left(F_{U T_{TT},T}^{\cos(2\phi_h-2\phi_{T_T })} + \epsilon  
F_{U T_{TT},L}^{\cos(2\phi_h-2\phi_{T_T})} 
\right)
\nonumber\\
& \hspace{2em}
+ \epsilon \cos2\phi_{T_T} F_{U T_{TT}}^{\cos2\phi_{T_T}}
\nonumber\\
& \hspace{2em}
+ \epsilon \cos(4\phi_h-2\phi_{T_T}) F_{U T_{TT}}^{\cos(4\phi_h-2\phi_{T_T})}
\nonumber \\
& \hspace{2em}
+ \sqrt{2\epsilon(1+\epsilon)} \cos(\phi_h-2\phi_{T_T}) 
F_{U T_{TT}}^{\cos(\phi_h-2\phi_{T_T})}
\nonumber \\
& \hspace{2em}
+ \left. \sqrt{2\epsilon(1+\epsilon)} \cos(3\phi_h-2\phi_{T_T}) 
F_{U T_{TT}}^{\cos(3\phi_h-2\phi_{T_T})} \right]
\nonumber \\
&+ T_{TT} (2\lambda_e) 
\left[\sqrt{1-\epsilon^2}\sin(2\phi_h-2\phi_{T_T})
F_{L T_{TT}}^{\sin(2\phi_h-2\phi_{T_T})} \right.
\nonumber\\
& \hspace{2em}
+ \sqrt{2\epsilon(1-\epsilon)}\sin(\phi_h-2\phi_{T_T}) 
F_{L T_{TT}}^{\sin(\phi_h-2\phi_{T_T})}
\nonumber \\
& \hspace{2em}
+ \left. \sqrt{2\epsilon(1-\epsilon)}\sin(3\phi_h-2\phi_{T_T}) 
F_{L T_{TT}}^{\sin(3\phi_h-2\phi_{T_T})} \right]\,.
\label{cross_section_tensor}
\end{align}
The decomposition predicts the dependence of the cross section on the beam and target
polarization and on the azimuthal angle of the produced hadron.
One observes:

\begin{enumerate}[(i)]
\item The form of the unpolarized and vector-polarized parts of the cross section for the
spin-1 target is identical to that for the spin-1/2 target \cite{Bacchetta:2006tn}.

\item The $\epsilon$ and $\phi_h$-dependence of the $T_{LL}$ tensor-polarized part
of the cross section (both for unpolarized and polarized lepton) is identical to that of
the unpolarized part.

\item The $\epsilon$ and $\phi_h$-dependence of the $T_{LT}$ tensor-polarized part
of the cross section is analogous to that of the $S_T$ vector-polarized part, only with
the replacement $\sin \to \cos$ because of the different parity of the spin vector and tensor,
and with the offset angle $\phi_S \to \phi_{T_L}$.

\item  The azimuthal dependence of the $T_{TT}$ tensor-polarized part of the cross section
is essentially different from that of the $S_T$ vector-polarized part. The first term in
the $T_{TT}$ cross section has a $\cos (2\phi_h - 2 \phi_{T_T})$ dependence, in the place
where the $S_T$ cross section has $\sin (\phi_h - \phi_{S})$. A term with
$\cos(4\phi_h-2\phi_{T_T})$ dependence appears in the $T_{TT}$ part of the cross section,
which has no counterpart in the $S_T$ part. The $4\phi_h$ dependence is absent in 
the spin-1/2 target and represents a unique feature of the spin-1 target.
\end{enumerate}

We recall that longitudinal and transverse ($L$ and $T$) target polarization here
are defined covariantly relative to the basis vectors $e_{L\ast}$ and $e_{x'}, e_{y'}$;
see Eq.~(\ref{eq:poltensor_LT}), which corresponds to the longitudinal and transverse
directions in the collinear frame. In the target rest frame this corresponds to target
polarization along the $\bm{q}$-vector direction, not along the lepton beam direction.

The decomposition also predicts the dependence of the cross section on $\epsilon$,
and thus the dependence on $y$ at fixed $x_A$ and $Q^2$, see Eq.~(\ref{epsilon_from_y}).
One observes that the terms proportional to the lepton helicity $\lambda_e$
involve factors $\sqrt{1 - \epsilon}$, which are $\propto y$ for $y \ll 1$.
This determines the dependence of beam spin asymmetries on the center-of-mass energy $s$ for
fixed $x_A$ and $Q^2$ (depolarization factors).

The structures in the SIDIS cross section can be classified according to their behavior under
(naive) time reversal \cite{Bacchetta:2006tn,Zhao:2025vol}. The spin-1 case can be treated in analogy 
to the spin-1/2 case. The structures with a single-spin dependence (on the lepton helicity $\lambda_e$,
or the target vector polarization $S_L$ and $S_T$) have a $\sin$-type dependence on $\phi_h$ or $\phi_S$
and are $T$-odd; the structures with no spin dependence or double-spin dependence have a
$\cos$-type dependence on the angles and are $T$-even. Note that only the vector polarization
of the target counts as ``spin dependence'' for this purpose, not the tensor polarization;
(see also the discussion in Appendix~\ref{app:density_matrix}). The $T$-odd structures in the
unpolarized and vector-polarized spin-1 cross section are the same as in the spin-1/2 cross section.
In the tensor-polarized spin-1 cross section, new $T$-odd structures appear in the
$\lambda_e$-dependent terms:
\begin{align}
& F_{L T_{LL}}^{\sin\phi_h},
\nonumber \\
& F_{L T_{LT}}^{\sin(\phi_h-\phi_{T_L})},
F_{L T_{LT}}^{\sin\phi_{T_L}}, F_{L T_{LT}}^{\sin(2\phi_h-\phi_{T_L})},
\nonumber \\
& F_{L T_{TT}}^{\sin(2\phi_h-2\phi_{T_T})}, F_{L T_{TT}}^{\sin(\phi_h-2\phi_{T_T})},
F_{L T_{TT}}^{\sin(3\phi_h-2\phi_{T_T})}.
\end{align}
The $T$-odd structures are special in regard to dynamics, as they require production amplitudes
with nonzero phases and are sensitive to final-state interactions (see {\secondpart}).

The analysis here assumes zero lepton mass. At this level the cross section depends only on the
lepton helicity, which is conserved in the scattering process. When the finite lepton mass is
included, transverse lepton polarization becomes possible, and the cross section develops single
and double-spin dependent structures involving the transverse lepton spin.

The general decomposition of the spin-1 SIDIS cross section obtained here agrees with the
one reported in Ref.~\cite{Zhao:2025vol}. That work uses a different convention for the
normalization of the tensor polarization intensity parameters; the correspondence with our convention
is explained in the article (our convention here is the same as in our earlier work \cite{Cosyn:2020kwu}).
The conventions for the tensor polarization angles are the same as ours.
The number of 41 independent structures also agrees with earlier results for the exclusive
deuteron electrodisintegration cross section \cite{Leidemann:1991qs}.

\subsection{Connection with inclusive scattering}
\label{sec:inclusive}
The structures in the semi-inclusive cross section arise from the interplay of the
initial-state momentum and polarization variables with the final-state hadron momentum.
When integrating over the final-state hadron momentum this information disappears,
and the allowed structures reduce to those of the inclusive cross section. We now establish the
connection of our semi-inclusive cross section with the
standard inclusive cross section parametrizations for the spin-1 target \cite{Hoodbhoy:1988am}.

We consider here the semi-inclusive cross section for production of a specific hadron $h$,
$e + A \rightarrow e' + h + X$, integrated over the momentum of $h$. We do not address the
question how the inclusive cross section $e + A \rightarrow e' + X$ could be obtained
from the semi-inclusive cross sections for a set of final-state hadrons by summing over the hadron species.
(This question requires additional assumptions about dynamics, e.g. in the context of the
quark fragmentation mechanism of hadron production.) In this sense the ``inclusive''
structure functions discussed here should be understood as ``integrated semi-inclusive''
structure functions, not as truly inclusive structure functions.

When integrating the semi-inclusive cross section Eq.~(\ref{cross_section_final})
over the observed hadron momentum, all terms with a harmonic azimuthal dependence
on $\phi_h$ drop out because the sine and cosine functions integrate to zero over the
full period $[0, 2\pi]$. In the remaining terms, the integrated semi-inclusive
structure functions can be matched with the conventional inclusive structure functions.
For the unpolarized and vector-polarized structures we obtain
\begin{subequations}
\begin{alignat}{2}
& \int d\Gamma_{P_h}  F_{UU,L} && = (1+\gamma^2)\frac{F_2}{x_A}-2 F_1,
\\
&  \int d\Gamma_{P_h} F_{UU,T}  && = 2F_1,
\\
&  \int d\Gamma_{P_h} F_{LS_L} && = 2(g_1-\gamma^2 g_2),
\\
&  \int d\Gamma_{P_h} F_{LS_T}^{\cos\phi_S} && = -2\gamma (g_1+g_2),
\end{alignat}
\end{subequations}
where $F_{1,2}$ and $g_{1,2}$ are the inclusive structure functions in the standard
convention (we suppress the dependence on $x_A$ and $Q^2$ for brevity).
For the tensor-polarized structures we obtain
\begin{subequations}
\label{eq:inclusivestruc}
\begin{align}
&  \int d\Gamma_{P_h} F_{U T_{LL},L}
\nonumber \\
& = \frac{1}{x_A}\left[2(1+\gamma^2)x_Ab_1-(1+\gamma^2)^2
\left(\frac {1} {3}b_2 + b_3 + b_4\right) \right.
\nonumber \\ 
& \left. - (1+\gamma^2)\left(\frac{1}{3}b_2-b_4\right)
-\left(\frac{1}{3}b_2-b_3 \right) \right] \,,
\nonumber\\
&  \int d\Gamma_{P_h}  F_{U T_{LL},T}
\nonumber\\
& = -\left[2(1+\gamma^2)b_1-\frac{\gamma^2}{x_A}\left(\frac{1}{6}
b_2-\frac{1}{2}b_3\right)\right ] \,,
\nonumber\\
&  \int d\Gamma_{P_h}  F_{U T_{LT}}^{\cos\phi_{ T_L}}
\nonumber \\
&= -\frac{\gamma}{2x_A}\left[(1+\gamma^2)\left(\frac{1}{3}
b_2-b_4\right)+\left(\frac{2}{3}b_2-2b_3\right)\right]\,,
\nonumber\\
& \int d\Gamma_{P_h}  F_{U T_{TT}}^{\cos2\phi_{ T_T}} 
  = - \frac{\gamma^2}{x_A}
\left(\frac{1}{6}b_2-\frac{1}{2}b_3 \right)\,.
\end{align}
\end{subequations}
where $b_{1-4}$ are the structure functions of the inclusive cross section
for a tensor-polarized spin-1 target in the convention of Ref.~\cite{Hoodbhoy:1988am}.
For reference we give here the relations between the tensors defined in 
Ref.~\cite{Hoodbhoy:1988am} and the tensors $\mathcal{A}_i^{\mu\nu}$
used in our decomposition (note that Ref.~\cite{Hoodbhoy:1988am} uses $\epsilon_{0123}=1$):
\begin{subequations}
\begin{align}
r^{\mu\nu} &=-2(1+\gamma^2) T_{LL}
(\mathcal { A } ^ { \mu\nu }_L-\mathcal {A}^{\mu\nu}_T),
\\
s^{\mu\nu} &= -2(1+\gamma^2)^2 T_{LL}\frac{1}{x_A}\mathcal{A}^{\mu\nu}_L,
\\
t^{\mu\nu} &= -(1+\gamma^2)\frac{1}{x_A}
\left\{ 2 T_{LL}\mathcal{A}^{\mu\nu}_L  
\right.
\nonumber \\
& -\gamma T_{LT}\left[\cos(\phi_h-\phi_{T_L}) \mathcal{A}^{\mu\nu}_{LT} \right.
\nonumber \\[1ex]
& \left. \left. +\sin(\phi_h-\phi_{T_L}) \mathcal{A}^{\mu\nu}_{LT''}\right]\right\},
\\
u^{\mu\nu} &= -\frac{1}{x_A}\left\{ 2 T_{LL}\left(\mathcal{A}^{\mu\nu}_L 
- \gamma^2 \mathcal{A}^{\mu\nu}_{TT} \right) \right.
\nonumber \\
& -2\gamma T_{LT} \left[ \cos(\phi_h-\phi_{T_L}) \mathcal{A}^{\mu\nu}_{LT}
+\sin(\phi_h-\phi_{T_L}) \mathcal{A}^{\mu\nu}_{LT''}\right]
\nonumber\\[1ex]
& +\gamma^2 T_{TT}\left[ \cos(2\phi_h-2\phi_{T_T}) 
\mathcal{A}^{\mu\nu}_{TT}
\right.
\nonumber \\[1ex]
& \left. \left. + \sin(2\phi_h-2\phi_{T_T}) \mathcal{A}^{\mu\nu}_{TT'}\right]  
\right\}.
\label{eq:hoodbhoy}
\end{align}
\end{subequations}

A special situation arises in the $T$-odd structures without $\phi_h$ dependence in
the semi-inclusive cross section. Time reversal invariance, combined with the hermiticity
of the electromagnetic current operator, forbids $T$-odd structures in the inclusive
cross section in one-photon exchange approximation \cite{Christ:1966zz}.
This implies that the integral of the corresponding semi-inclusive structure functions
must be zero (see Ref.~\cite{Diehl:2005pc} for the spin-1/2 case)
\begin{subequations}
\label{integrated_todd}
\begin{align}
\int d\Gamma_{P_h} \; F_{US_T}^{\sin\phi_S}
&= 0,
\\
\int d\Gamma_{P_h} \; F_{L T_{LT}}^{\sin\phi_{T_L}} 
&= 0.
\end{align}
\end{subequations}
Here the vanishing of the structures in the inclusive cross section is not realized by
averaging out of the $\phi_p$ dependence (the terms in the cross section have no dependence
on $\phi_h$) but by a dynamical constraint on the structure functions, involving the
dependence on the longitudinal and transverse hadron momentum. Note that the
relations Eq.~(\ref{integrated_todd}) are limited to the one-photon exchange approximation;
at two-photon exchange level the inclusive cross section develops a transverse
single-spin dependence \cite{Afanasev:2007ii}.

\section{Polarization observables}
\label{sec:observables}
\subsection{Preparation of target polarization}
\label{subsec:preparation}
The decomposition of the semi-inclusive differential cross section implies a definite dependence
on the kinematic variables.
The dependence on the beam and target polarization, the virtual photon polarization, and the
final-state hadron azimuthal angle $\phi_h$ is kinematic and shown explicitly in the expressions
in Sec.~\ref{subsec:complete}. The dependence on the DIS variables
$x_A$ and $Q^2$, and the final-state hadron longitudinal and transverse momenta, $z_h$ and $|\bm{P}_{hT}|$,
is dynamical and contained in the invariant structure functions, which represent the basic independent
structures characterizing the process. Observables for a defined experimental setup can be derived from
the complete expressions of the differential cross section. Depending on the setup and the goals of
the measurement, a large variety of observables (polarized differential cross sections, azimuthal harmonics,
spin asymmetries etc.) can be constructed and used to separate the independent structures.
We now describe some basic techniques for constructing the observables and separating the structures.
Specific examples will be presented for DIS on the deuteron with spectator nucleon tagging in {\secondpart}.

We first describe the preparation of the target polarization of the spin-1 target.
We consider a setup where the spin-1 target is polarized along a fixed spatial direction in
a given frame (to be specified); measurements are performed in the pure spin states with projection
$\Lambda = (-1, 0, +1)$ along the axis; and then certain sums or differences of cross-section measurements in
the pure spin states are taken. This scheme covers most actual experimental setups; a general setup
can be modeled by superposing several such measurements. The expressions are presented in
relativistically covariant form, using the covariant representation of the spin density matrix
in Sec.~\ref{sec:densitymatrix}, and can be applied to fixed-target and colliding-beam experiments.

The polarization axis is described covariantly by a spacelike 4-vector $N$ with
\begin{align}
(NP) = 0, \hspace{1em} N^2 = -1.
\end{align}
The spin density matrix of the pure spin state with spin projection $\Lambda$
on the axis $N$ is given by the general expression Eq.~(\ref{density_tensor_parametrization})
with the parameters \cite{Cosyn:2020kwu}
\begin{subequations}
\label{pure_state_density}
\begin{align}
S^\alpha &= \Lambda N^\alpha,
\\
T^{\alpha\beta} &= \frac{1}{6} W(\Lambda)
\left( g^{\alpha\beta} - \frac{P^\alpha P^\beta}{P^2} + 3 N^\alpha N^\beta \right),
\\
W(\Lambda) &\equiv \textrm{$(1,-2,1)$ for $\Lambda = (+1,0,-1)$}.
\end{align}
\end{subequations}
The 4-vector $N$ then can be expanded in the basis vectors of Set II, Eq.~(\ref{basis_longit_set2}), as
\begin{subequations}
\label{N_expanded}
\begin{align}
N =& - \cos\theta_N \, e_{L\ast}
\nonumber \\[1ex]
& + \sin\theta_N [\cos \phi_N \, e_{x} - \sin \phi_N \, e_{y}]
\\[1ex]
=& - \cos\theta_N \, e_{L\ast}
\nonumber \\[1ex]
& + \sin\theta_N [\cos (\phi_h - \phi_N) \, e_{x'} + \sin (\phi_h - \phi_N) \, e_{y'}] .
\end{align}
\end{subequations}
Here the coefficients are expressed in terms of two angular parameters, $\theta_N$ and $\phi_N$.
$\theta_N$ is defined such that, in the target rest frame, where $N = (0, \bm{N})$ and
the basis vectors are given by Eqs.~(\ref{set2_collinear_frame}), (\ref{collinear_frame_lepton}),
and (\ref{collinear_frame_hadron}),
$\theta_N$ is the polar angle of $\bm{N}$ relative to the $z$-axis.
The azimuthal angle $\phi_N$ is defined as specified for $\phi_S$ in Fig.~\ref{fig:kin}.
The invariant polarization parameters generated by the pure state are then
computed by contracting Eq.~(\ref{pure_state_density}) with the basis vectors,
as specified in Eq.~(\ref{eq:poltensor_LT}). We obtain
\begin{subequations}
\label{pure_state_invariant}
\begin{align}
S_L &= \Lambda \cos \theta_N,
\\
S_T &= \Lambda \sin\theta_N,
\hspace{1em}
\phi_S = \phi_N, 
\\
T_{LL} &= \tfrac{1}{6} W(\Lambda) (-1 + 3 \cos^2 \theta_N)
\\
T_{LT} &= \tfrac{1}{2} W(\Lambda) \sin \theta_N \cos \theta_N,
\hspace{1em}
\phi_{T_L}  = \phi_N,
\\
T_{TT} &= \tfrac{1}{2} W(\Lambda) \sin^2 \theta_N,
\hspace{1em}
\phi_{T_T}  = \phi_N.
\end{align}
\end{subequations}
One observes that all the three vector polarization parameters $\{S_L, S_T, \phi_S\}$
and the five tensor polarization parameters $\{T_{LL}, T_{LT}, T_{TT}, \phi_{T_L}, \phi_{T_T} \}$
can be assigned nontrivial values through suitable choices of the pure state polarization
parameters $\{\Lambda, \theta_N, \phi_N \}$.

In most experiments the target polarization is prepared along an axis defined relative
to the lepton beam direction (parallel or perpendicular to it).
This includes (i) fixed-target experiments, where the lepton beam defines an axis in the target rest frame;
(ii) colliding-beam experiments with head-on collisions of the lepton and nuclear beams
(zero crossing angle); this situation is equivalent to fixed-target when viewed in target rest frame.

%
%
\begin{figure}[t]
\begin{center}
\includegraphics[width=0.95\columnwidth]{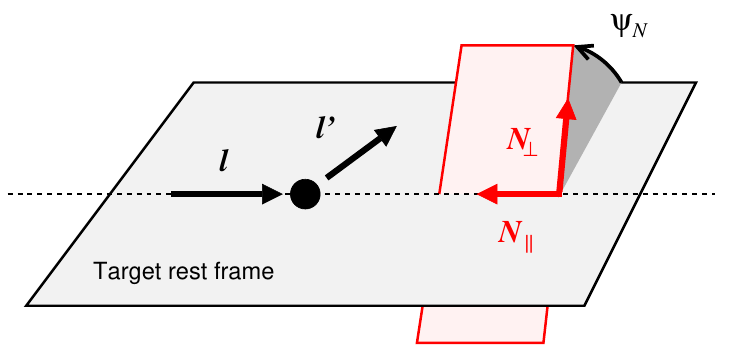}
\end{center}
\caption{
\label{fig:target_rest_frame}
Target polarization relative to the lepton beam axis in the target rest frame.
Shown are the spatial components of the vector $\bm{N}$ in the target rest frame.
For parallel polarization ($\parallel$), $\bm{N}$ points in the direction opposite to the initial lepton momentum.
For perpendicular polarization ($\perp$), the angle $\psi_N$ is measured from the plane
spanned by the final lepton momentum in the sense as indicated.}
\end{figure}
For target polarization parallel to the lepton beam axis (denoted by $\parallel$)
the axis 4-vector is given by
\begin{align}
& N = \frac{-l + (e_P l) e_P}{\sqrt{(e_P l)^2}},
\label{N_parallel}
\end{align}
which in the target rest frame becomes the unit vector along the negative lepton momentum,
\begin{align}
N = (0, -\bm{l}/|\bm{l}|).
\end{align}
The polarization is thus measured relative to the lepton beam axis in the target rest frame,
with $\Lambda = +1 (-1)$ corresponding to polarization opposite (along) the lepton momentum
(see Fig.~\ref{fig:target_rest_frame}).
For this choice of polarization axis the coefficients in Eq.~(\ref{N_expanded}) are
\begin{subequations}
\begin{align}
\cos\theta_N &= \frac{1 + \gamma^2 y/2}{\sqrt{1 + \gamma^2}}
\\
\sin\theta_N &= \frac{\gamma \sqrt{1 - y - \gamma^2 y^2/4}}{\sqrt{1 + \gamma^2}},
\hspace{1em}
\phi_N = 0.
\end{align}
\end{subequations}
For target polarization perpendicular to lepton beam axis (denoted by $\perp$)
the axis 4-vector is given by
\begin{subequations}
\begin{align}
&N = \cos\psi_N \, e_\perp - \sin\psi_N \, e_y,
\label{N_perp}
\\[1ex]
&e_\perp^\alpha \equiv \frac{\epsilon^{\alpha\beta\gamma\delta}
e_{P, \beta} e_{y, \gamma} l_\delta}{\sqrt{(e_P l)^2}}
= \frac{(e_x l) e_{L\ast}^\alpha - (e_{L\ast} l) e_x^\alpha}{\sqrt{(e_x l)^2 + (e_{L\ast} l)^2}},
\\[1ex]
&(e_P e_\perp), (e_y e_\perp), (l e_\perp) = 0,
\end{align}
\end{subequations}
which in the target rest frame becomes
\begin{subequations}
\begin{align}
& N = (0, \bm{N}),
\hspace{1em}
\bm{N} = \cos\psi_N \bm{e}_\perp - \sin\psi_N \bm{e}_y,
\\
& \bm{e}_\perp \equiv -\bm{e}_y \times \frac{\bm{l}}{|\bm l|}
= \frac{l^x \bm{e}_z - l^z \bm{e}_x }{|\bm l|}.
\end{align}
\end{subequations}
In the last expression, $\bm{e}_z$ and $\bm{e}_x$ are the 3-dimensional components of the basis vectors
in the target rest frame as defined in Sec.~\ref{subsec:basis_vectors}, and $l^z$ and $l^x$ are the
components of $\bm{l}$ along these vectors.
$\bm{e}_\perp$ is the unit vector in the lepton scattering plane perpendicular to $\bm{l}$, such that
$\{\bm{e}_\perp, \bm{e}_y, -\bm{l}/|\bm{l}|\}$ form a right-handed system (see Fig.~\ref{fig:target_rest_frame}).
For this choice of polarization axis the coefficients in Eq.~(\ref{N_expanded}) are
\begin{subequations}
\begin{align}
\cos\theta_N &= \frac{\gamma\sqrt{1-y-\gamma^2y^2/4}}{\sqrt{1+\gamma^2}} \, \cos \psi_N ,
\\[1ex]
\sin\theta_N \cos\phi_N &= \frac{1+\gamma^2y/2}{\sqrt{1+\gamma^2}}\, \cos\psi_N ,
\\[2ex]
\sin\theta_N \sin\phi_N &= \sin\psi_N .
\end{align}
\end{subequations}
Note that the definition of the angle $\psi_N$ in Eq.~(\ref{N_perp}) and Fig.~\ref{fig:target_rest_frame}
differs from Ref.~\cite{Cosyn:2020kwu} (see Appendix~\ref{app:correspondence}).
Using the above expressions one can easily compute the invariant polarization parameters
for a given experimental setup in fixed-target or colliding-beam experiments.

\subsection{Separation of spin structures}
Equation~(\ref{pure_state_invariant}) gives the values of the invariant polarization parameters
in pure spin states with polarization axis $N$ and spin projection $\Lambda$.
The parts of the cross section proportional to these parameters then can be separated by
performing measurements in pure spin states with different $\Lambda$ and taking sums/differences.
Specifically:

(i) The unpolarized part of the cross section is obtained by performing measurements
in all three spin states and summing them,
\begin{align}
&\textrm{$U$: \hspace{1em} $\sigma(\Lambda = +1) + \sigma(\Lambda = -1) + \sigma(\Lambda = 0)$.}
\end{align}
The summation over all three states sets both vector and tensor polarization to zero, because
$W(+1) + W(0) + W(-1) = 0$. Note that the summation over only the $\Lambda = \pm 1$ states
has nonzero tensor polarization, because $W(+1) + W(-1) = 2 \neq 0$. The $\Lambda = 0$ state
is thus needed for unpolarized measurements.

(ii) The vector-polarized part of the cross section is isolated by
taking the difference of measurements in spin states with $\Lambda = \pm 1$,
\begin{align}
&\textrm{$S$: \hspace{1em} $\sigma(\Lambda = +1) - \sigma(\Lambda = -1)$.}
\end{align}
In the difference the values of $S_L$ and $S_T$ add up,
while the values of $T_{LL}, T_{LT}, T_{TT}$ sum to zero, see Eq.~(\ref{pure_state_invariant}).
The $\Lambda = 0$ state is not needed for measurements of vector-polarized structures.
Longitudinal and transverse polarization can be selected by choosing $\theta_N = 0$ and $\pi/2$,
respectively.

(iii) The tensor-polarized part of the cross section is isolated by
performing measurements in all three spin states and taking the combination
\begin{align}
&\textrm{$T$: \hspace{1em} $\sigma(\Lambda = +1) + \sigma(\Lambda = -1) - 2\, \sigma(\Lambda = 0)$.}
\end{align}
The combination removes the unpolarized and vector-polarized terms. 
The values of $\{ T_{LL}, T_{LT}, T_{TT} \}$ can be assigned by choosing appropriate
angles of the pure-state polarization axis, see Eq.~(\ref{pure_state_invariant}).
Special choices are
\begin{align}
\label{tensor_angles}
\begin{array}{r|rrr}
\cos\theta_N & T_{LL} & T_{LT} & T_{TT} \\
\hline
\pm 1  & \checkmark & 0 & 0 \\
\pm \frac{1}{\sqrt{3}}  & 0 & \checkmark & \checkmark \\
0  & \checkmark & 0 & \checkmark \\
\end{array}
\end{align}
Using suitable combinations of measurements with these special axis angles
one can separate the various tensor-polarized structures.

The preparation of the spin density matrix of the spin-1 target as a superposition of
pure states quantized along certain directions practiced here is analogous to the
preparation of photon polarization in optics \cite{Berestetskii:1982qgu}.
The polarization parameters can be associated with probability differences
(intensity differences) measured along certain polarization directions.
Measurements along several directions, including relative angles other
than 90$^\circ$, are needed to completely specify the polarization of the spin-1 system;
see Eq.~(\ref{tensor_angles}).

\subsection{Azimuthal angle dependence}
The dependence of the differential cross section on the hadron azimuthal angle $\phi_h$ and the polarization
angles $\phi_S, \phi_{T_L}, \phi_{T_T}$ is explicit and described by simple trigonometric functions, see
Eqs.~(\ref{cross_section_unpol}) -- (\ref{cross_section_tensor}). The individual structure functions
can be extracted by measuring the azimuthal dependence of the differential cross section and computing
certain averages weighted with trigonometric functions. We use the notation
\begin{subequations}
\begin{align}
\langle \mathcal{F} \rangle_h
&\equiv \frac{\displaystyle \int_0^{2\pi}
\frac{d\phi_h}{2 \pi} \;
h(\phi_h) \; \mathcal{F} }
{\displaystyle \int_0^{2\pi} \frac{d\phi_h}{2 \pi} \; h^2(\phi_h)},
\\[2ex]
h(\phi_h, \phi_{\rm pol}) &\equiv 1, \cos (n \phi_h), 
\sin (n \phi_h) \hspace{1em} \text{($n =$ 1--4).}
\end{align}
\end{subequations}
The averages can be computed by using the addition theorems of the trigonometric functions to
separate the dependencies on $\phi_h$ and $\phi_S, \phi_{T_L}, \phi_{T_T}$
in the various terms in Eqs.~(\ref{cross_section_unpol})--(\ref{cross_section_tensor}),
and using the orthogonality properties of the elementary trigonometric functions in $n\phi_h$.
The $\phi_h$-independent differential cross section (average with $h = 1$) is given by
Eq.~(\ref{cross_section_final}) with 
\begin{align}
\langle \mathcal{F} \rangle_1
&= \int_0^{2\pi} \frac{d\phi_h}{2 \pi} \; \mathcal{F}
\nonumber \\
&= F_{UU,T} + \epsilon F_{UU,L}
\nonumber \\
&+S_L (2\lambda_e) \sqrt{1-\epsilon^2}F_{LS_L}
\nonumber \\
&+S_T (2\lambda_e) \sqrt{2\epsilon(1-\epsilon)} \cos\phi_S F_{LS_T}^{\cos\phi_S} 
\nonumber \\
&+ T_{LL} \left( F_{U T_{LL},T} + \epsilon F_{U T_{LL},L} \right)
\nonumber \\
&+ T_{LT} \sqrt{2\epsilon(1+\epsilon)} \cos\phi_{T_L} F_{U T_{LT}}^{\cos\phi_{T_L}}
\nonumber \\
&+ T_{LT} (2\lambda_e) \sqrt{2\epsilon(1-\epsilon)} \sin\phi_{T_L} F_{L T_{LT}}^{\sin\phi_{T_L}}
\nonumber \\
&+T_{TT} \epsilon \cos2\phi_{T_T} F_{U T_{TT}}^{\cos2\phi_{T_T}} .
\label{cross_section_harmonic_1}
\end{align}
The averages with $n =$ 0, 1 and 2 generally have terms with all the spin dependencies
present in the cross section, because spin-orbit effects in the SIDIS process mix the
azimuthal and spin dependence of the structures.

The averages with $n = 3$ and 4 have fewer terms and are of particular interest.
The $n = 3$ averages are
\begin{subequations}
\begin{align}
&\langle \mathcal{F} \rangle_{\cos 3\phi_p}
\nonumber \\
&= -S_T \epsilon \sin\phi_S F_{US_T}^{\sin(3\phi_h-\phi_S)}
\nonumber \\
&+ T_{LT} \epsilon \cos (\phi_{T_L}) F_{U T_{LT}}^{\cos(3\phi_h-\phi_{T_L})}
\nonumber \\
&+ T_{TT} \sqrt{2\epsilon(1+\epsilon)} \cos (2 \phi_{T_T}) 
F_{U T_{TT}}^{\cos(3\phi_h-2\phi_{T_T})}
\nonumber \\
&-T_{TT} (2\lambda_e)
\sqrt{2\epsilon(1-\epsilon)} \sin(2\phi_{T_T}) 
F_{L T_{TT}}^{\sin(3\phi_h-2\phi_{T_T})}, 
\\[2ex]
&\langle \mathcal{F} \rangle_{\sin 3\phi_p}
\nonumber \\
&= S_T \epsilon \cos\phi_S F_{US_T}^{\sin(3\phi_h-\phi_S)}
\nonumber \\
&+ T_{LT} \epsilon \sin (\phi_{T_L}) F_{U T_{LT}}^{\cos(3\phi_h-\phi_{T_L})}
\nonumber \\
&+ T_{TT} \sqrt{2\epsilon(1+\epsilon)} \sin(2\phi_{T_T}) 
F_{U T_{TT}}^{\cos(3\phi_h-2\phi_{T_T})}
\nonumber \\
&+ T_{TT} (2\lambda_e)
\sqrt{2\epsilon(1-\epsilon)} \cos (2\phi_{T_T}) 
F_{L T_{TT}}^{\sin(3\phi_h-2\phi_{T_T})} .
\end{align}
\end{subequations}
They contain terms with transverse vector polarization, $S_T$, and with mixed and
transverse tensor polarization, $T_{LT}$ and $T_{TT}$. The terms can be separated through their
dependence on the polarization angles $\phi_S$ and $\phi_{T_L}, \phi_{T_T}$.
For example,
\begin{subequations}
\begin{align}
&\langle \mathcal{F} \rangle_{\cos 3\phi_p} (\phi_S = 0, \phi_{T_L} = \phi_{T_T} = 0)
\nonumber \\
&= T_{LT} \epsilon F_{U T_{LT}}^{\cos(3\phi_h-\phi_{T_L})}
\nonumber \\
&+ T_{TT} \sqrt{2\epsilon(1+\epsilon)} 
F_{U T_{TT}}^{\cos(3\phi_h-2\phi_{T_T})},
\\[2ex]
&\langle \mathcal{F} \rangle_{\sin 3\phi_p} (\phi_S = \pi/2, \phi_{T_L} = \phi_{T_T} = \pi/2)
\nonumber \\
&= T_{LT} \epsilon F_{U T_{LT}}^{\cos(3\phi_h-\phi_{T_L})}
\nonumber \\
&- T_{TT} (2\lambda_e)
\sqrt{2\epsilon(1-\epsilon)} 
F_{L T_{TT}}^{\sin(3\phi_h-2\phi_{T_T})},
\end{align}
\end{subequations}
which contain only tensor-polarized structures. Such measurements can be performed in pure target
spin states polarized relative to the lepton beam axis, in which $\phi_S = \phi_{T_L} = \phi_{T_T} = \phi_N$;
see Eq.~(\ref{pure_state_invariant}).
The $n = 4$ averages are
\begin{subequations}
\begin{align}
&\langle \mathcal{F} \rangle_{\cos 4\phi_p}
= T_{TT} \epsilon \cos (2\phi_{T_T}) F_{U T_{TT}}^{\cos(4\phi_h-2\phi_{T_T})},
\\[2ex]
&\langle \mathcal{F} \rangle_{\sin 4\phi_p}
= T_{TT} \epsilon \sin (2\phi_{T_T}) F_{U T_{TT}}^{\cos(4\phi_h-2\phi_{T_T})}.
\end{align}
\end{subequations}
which contain only the transverse tensor-polarized term. The $n = 4$ azimuthal harmonics
represent unique features of the spin-1 target and are absent for spin-1/2.

By computing the azimuthal harmonics and choosing appropriate target polarization axes,
one can separate vector- and tensor-polarized structures at the cross section level.
Note that, in the terms of the cross section where structure functions with $T$ and $L$
photon polarization appear with the same spin and azimuthal angle dependence,
the structure functions can only be separated through the dependence on $\epsilon$ (Rosenbluth separation).

The discussion here assumes that the azimuthal harmonics of the cross section are separated
by integrating the angular dependence with appropriate weights. In actual experiments the harmonics
can also be extracted through fits of the angular dependence, with possibly incomplete angular
coverage, and multidimensional fits in several kinematic variables.

\subsection{Spin asymmetries}
\label{subsec:spin_asymmetries}
Spin asymmetries are ratios of differences and sums of the cross sections in the
lepton and target spin variables and are standard observables in the analysis of SIDIS.
On the experimental side, the luminosity and detection efficiency cancel in the ratios,
reducing the systematic uncertainty.
On the theoretical side, the flux factors in the cross section cancel, reducing the
kinematic dependence and giving direct access to the structure functions.
The theoretical expressions of the asymmetries can generally be represented in the form
\begin{align}
A = & \sum \textrm{(kinematic factors)}
\nonumber \\
& \times \textrm{(structure function ratios)},
\end{align}
where the kinematic factors (depolarization factors) depend on $\epsilon$ and on
the invariant polarization parameters, and the structure function ratios contain
the dynamical information on the target structure
(depending on the details, these ratios may have additional kinematic dependencies).
With the complex dependence of the spin-1 SIDIS cross section on the target polarization and the
final-state hadron momentum, a large variety of asymmetries can be formed and used for experimental analysis.
The theoretical expressions can be derived from the formulas of the differential cross section,
Eqs.~(\ref{cross_section_unpol})--(\ref{cross_section_tensor}), and the invariant polarization parameters
achieved in measurements with a given pure-state target polarization, Eq.~(\ref{pure_state_invariant}) et seq.
Here we describe only the basic asymmetries of the $\phi_p$-averaged differential cross section
for vector and tensor polarization, in order to connect with earlier results in Ref.~\cite{Cosyn:2020kwu}
and the experimental literature.

The vector-polarized asymmetry is defined as
\begin{subequations}
\label{double_spin_asymmetry}
\begin{align}
& A^V \equiv
\frac{d\sigma (++) - d\sigma (-+) - d\sigma (+-) + d\sigma (--)}
{d\sigma (++) + d\sigma (-+) + d\sigma (+-) + d\sigma (--)},
\\[1ex]
&d\sigma (\pm\pm) \equiv d\sigma (\lambda_e = \pm\tfrac{1}{2}, \Lambda = \pm 1).
\end{align}
\end{subequations}
Here the differential cross section is assumed to be averaged over $\phi_p$
[given by Eq.~(\ref{cross_section_final}) with $\mathcal{F} \rightarrow \langle \mathcal{F} \rangle_1$,
Eq.~(\ref{cross_section_harmonic_1})], and differential in the remaining variables of the hadron
phase space element. The differential cross section depends on the lepton helicity $\lambda_e$
and the deuteron spin projection $\Lambda$ along the quantization axis $N$; see Sec.~\ref{subsec:preparation}.
The difference of spin-dependent cross sections in the numerator isolates the vector-polarized
structures in the deuteron; the sum in the denominator involves the unpolarized and tensor-polarized structures.
The asymmetry takes values $A^V \in [-1,1]$. The asymmetries for target polarization parallel and perpendicular
to the lepton beam are denoted by $\parallel$ and $\perp$ (see Sec.~\ref{subsec:preparation}).
For target polarization parallel to the lepton beam, Eq.~(\ref{N_parallel}) et seq., the asymmetry becomes
\begin{subequations}
\label{asymmetry_parallel_expanded}
\begin{align}
A^V_{\parallel} &= D_{\parallel [S_L]} \frac{F_{L S_L}}{\Sigma_\parallel F}
+ D_{\parallel [S_T]} \frac{F_{L S_T}^{\cos\phi_S} }{\Sigma_\parallel F},
\\[1ex]
\Sigma_\parallel F &\equiv F_{UU,T} + \epsilon F_{UU,L} 
\nonumber \\[1ex]
&+ D_{\parallel [T_{LL}]} \, (F_{UT_{LL},T} + \epsilon F_{UT_{LL},L})
\nonumber \\[1ex]
&+ D_{\parallel [UT_{LT}]} \, F_{UT_{LT}}^{\cos \phi_{T_L}} + D_{\parallel [T_{TT}]}F_{UT_{TT}}^{\cos 2\phi_{T_T}},
\end{align}
\end{subequations}
where the kinematic factors $D_{\parallel [S_L]}, ... D_{\parallel [T_{TT}]}$ are
given in Ref.~\cite{Cosyn:2020kwu}, Eqs.~(3.71) and (3.65).
For target polarization perpendicular to the lepton beam, Eq.~(\ref{N_perp}) et seq., the
asymmetry becomes
\begin{subequations}
\label{asymmetry_perp_expanded}
\begin{align}
A^V_{\perp} &= D_{\perp [S_L]} \frac{F_{L S_L}}{\Sigma_\perp F}
+ D_{\perp [S_T]} \frac{F_{L S_T}^{\cos\phi_S} }{\Sigma_\perp F},
\\[1ex]
\Sigma_\perp F &\equiv F_{UU,T} + \epsilon F_{UU,L} 
\nonumber \\[1ex]
&+ D_{\perp [T_{LL}]} \, (F_{UT_{LL},T} + \epsilon F_{UT_{LL},L})
\nonumber \\[1ex]
&+ D_{\perp [UT_{LT}]} \, F_{UT_{LT}}^{\cos \phi_{T_L}} + D_{\perp [T_{TT}]}F_{UT_{TT}}^{\cos 2\phi_{T_T}},
\end{align}
\end{subequations}
where the kinematic factors $D_{\perp [S_L]}, ...D_{\perp [T_{TT}]}$ are
given in Ref.~\cite{Cosyn:2020kwu}, Eqs.~(3.72) and (3.66).

The vector polarized asymmetry Eq.~(\ref{double_spin_asymmetry}) is the ``two-state asymmetry,''
where the denominator is formed using only the $\Lambda = \pm 1$ states, but not the $\Lambda = 0$ state.
(In Ref.~\cite{Cosyn:2020kwu} this asymmetry is denoted by subscript '(2)', which we drop here for brevity.)
One can also consider the three-state asymmetry, where the denominator is the average over all three
spin states, $\Lambda = \pm 1$ and $0$, and does not involve tensor-polarized structures; the corresponding
kinematic factors are given in Ref.~\cite{Cosyn:2020kwu}.
The two-state asymmetry is more interesting because it is easier to measure and is strictly bounded by $[-1, 1]$.

The tensor-polarized asymmetry is defined as 
\begin{subequations}
\label{eq:AT_gen}
\begin{align}
& A^T \equiv
\frac{d\sigma (+) + d\sigma (-) - 2 d\sigma (0)}
{d\sigma (+) + d\sigma (-) + d\sigma (0)},
\\[1ex]
&d\sigma (\Lambda) \equiv \tfrac{1}{2} \sum_{\lambda_e} d\sigma (\lambda_e, \Lambda).
\end{align}
\end{subequations}
Here the differential cross section is $\phi_p$-averaged and averaged over the
lepton helicity, and depends on the target spin projection $\Lambda$
along the quantization axis $N$; see Sec.~\ref{subsec:preparation}.
The cross section difference in the numerator isolates the tensor-polarized deuteron structures;
the denominator includes only the unpolarized structures. Equation~(\ref{eq:AT_gen}) takes values
$A \in [-2,1]$. For target polarization parallel to the lepton beam, Eq.~(\ref{N_parallel}) et seq.,
the tensor-polarized asymmetry becomes
\begin{subequations}
\label{asymmetry_tensor_parallel_expanded}
\begin{align}
A^T_{\parallel} &=
2 D_{\parallel [T_{LL}]} \frac{F_{UT_{LL},T} + \epsilon F_{UT_{LL},L}}
{F_U}
\nonumber \\
&+ 2 D_{\parallel [UT_{LT}]} \frac{F_{UT_{LT}}^{\cos \phi_{T_L}}}
{F_U}
+ 2 D_{\parallel [T_{TT}]} \frac{F_{UT_{TT}}^{\cos 2\phi_{T_T}}}
{F_U},
\\[1ex]
F_U &\equiv F_{UU,T} + \epsilon F_{UU,L}.
\end{align}
\end{subequations}
The kinematic factors $D_{\parallel [T_{LL}]}, D_{\parallel [UT_{LT}]}, D_{\parallel [T_{TT}]}$
are given in Ref.~\cite{Cosyn:2020kwu}, Eq.~(3.65); the same factors appear in the
denominator of the vector-polarized asymmetry Eq.~(\ref{asymmetry_parallel_expanded}).
Likewise, for target polarization perpendicular to the lepton beam,
Eq.~(\ref{N_perp}) et seq., the tensor-polarized asymmetry becomes
\begin{subequations}
\label{asymmetry_tensor_perp_expanded}
\begin{align}
A^T_{\perp} &=
2 D_{\perp [T_{LL}]} \frac{F_{UT_{LL},T} + \epsilon F_{UT_{LL},L}}
{F_U}
\nonumber \\
&+ 2 D_{\perp [UT_{LT}]} \frac{F_{UT_{LT}}^{\cos \phi_{T_L}}}
{F_U}
+ 2 D_{\perp [T_{TT}]} \frac{F_{UT_{TT}}^{\cos 2\phi_{T_T}}}
{F_U},
\end{align}
\end{subequations}
where the denominator $F_U$ is the same as in Eq.~(\ref{asymmetry_tensor_parallel_expanded}).
The kinematic factors $D_{\perp [T_{LL}]}, D_{\perp [UT_{LT}]}, D_{\perp [T_{TT}]}$
are given in Ref.~\cite{Cosyn:2020kwu}, Eq.~(3.66).

The spin asymmetries considered here are for the $\phi_p$-averaged differential cross section.
More complex spin asymmetries can be formed from the $\phi_p$-dependent differential cross section,
where all the terms in the differential cross section contribute. The corresponding expressions
can be derived from the general decomposition of the cross section Eq.~(\ref{cross_section_final}).
In particular, when the $\phi_p$-dependence is included, one can form:
(i) lepton single-spin asymmetries, involving the $\lambda_e$ dependent term of $\mathcal{F}_U$;
(ii) vector-polarized target single-spin asymmetries, involving the $S_L$ and $S_T$ dependent
terms in $\mathcal{F}_S$;
(iii) tensor-polarized double-spin asymmetries, involving the $T_{LL} \lambda_e, T_{LT} \lambda_e$
and $T_{TT} \lambda_e$ dependent terms in $\mathcal{F}_T$.

\section{Summary}
\label{sec:summary}
In this work we have derived the general form of the cross section for SIDIS
on a polarized spin-1 target. The expressions obtained are general
and do not rely on assumptions about strong interaction dynamics (only $P$ and $T$ invariance).
The kinematic factors are presented in their exact form, and the cross section
decomposition can be used at all energies, in the deep-inelastic regime as well as in
low-energy hadron production or nuclear breakup processes.

Special attention has been paid to a relativistically covariant formulation of the target polarization
and the final-state hadron distributions. The orthonormal basis 4-vectors can be constructed
from the particle momenta in any reference frame where they are available, and the invariant
polarization parameters and final-state variables can be computed as scalar products
from the 4-vector components in that reference frame. The covariant spin density matrix of
the spin-1 target can be specified in any frame where the experimental polarization is available
(e.g.\ the target rest frame). This enables simple and efficient calculations and avoids the
use of explicit Lorentz transformations. 

At the same time, the covariant expressions can be evaluated and interpreted in the class of collinear frames,
where they have a simple interpretation. The class of collinear frames contains the target rest frame
as a special instance. The connection with the target rest frame is useful for fixed-target experiments,
where target polarization and the kinematic vectors are directly given in the target rest frame;
and for colliding-beam experiments, where the polarization of the beam particle has a simple
relation to that in the rest frame.

The form of the SIDIS cross section for the vector-polarized spin-1 target is the same as for the
polarized spin-1/2 target. The new feature of the spin-1 target is the presence of
tensor-polarized structures, which include lepton helicity-independent and dependent
structures and give rise to new azimuthal modulations, including unique $\cos (4\phi_h)$
and $\sin (4\phi_h)$ dependencies. Measurement of these azimuthal modulations could
demonstrate the presence of nonzero tensor polarization in the target without polarimetry.

The general form of the cross section obtained here can be applied to SIDIS on the polarized deuteron
in the current fragmentation region (identified hadron/jets). In the asymptotic regime the SIDIS
structure functions can be computed using QCD factorization and become convolutions of transverse
momentum-dependent (TMD) parton densities and fragmentation functions \cite{Boussarie:2023izj}.
Extensive efforts have been made in analyzing these structures for the spin-1/2 target;
the techniques can be extended to the spin-1 target, see Ref.~\cite{Zhao:2025vol}.

The results obtained here can also be applied to SIDIS on the polarized deuteron in target fragmentation region,
in particular the nuclear breakup process (spectator nucleon tagging). For this process the semi-inclusive
structure functions can be computed by separating nuclear and nucleon structure and using low-energy
nuclear dynamics to predict the nuclear momentum distributions and breakup amplitudes; see {\secondpart}.
Large tensor polarization effects can be achieved by selecting final states with spectator momenta
$\sim$ 300--400 MeV (in the deuteron rest frame), where the $D$-wave in the deuteron wave function
is dominant.

The structural analysis of the semi-inclusive cross section presented here can be extended from the
spin-1 target to higher-spin targets. The basic steps are outlined in Appendix~\ref{app:higher_spin}.

The present study describes tensor-polarized structures in the cross section of scattering from
a spin-1 target, supported by the polarization of the initial state. Tensor-polarized structures can also
appear in hadronic processes with spin 1/2 $\rightarrow$ 3/2 transitions, when formed from the spin
variables in the initial and final states. For example, a tensor-polarized parton density appears
in the $N \rightarrow \Delta$ transition and can be studied in the context of generalized
parton distributions \cite{Kim:2025ilc}. Tensor-polarized observables using final-state spin
variables could also be formed in SIDIS. One example would be SIDIS on the polarized nucleon
with detection of a $\Delta$ in the target fragmentation region, in which the
$\Delta \rightarrow \pi N$ decay distribution is coupled with the spin of the $\Delta$.
This process can be analyzed using QCD factorization with fracture functions and
deserves further study \cite{Trentadue:1993ka,Anselmino:2011ss}.

\appendix

\section{Correspondence with earlier work}
\label{app:correspondence}
In this appendix we summarize the changes in definitions relative to our earlier work \cite{Cosyn:2020kwu}.
The changes are caused by our use of the Trento convention for the azimuthal angles in the present work.

(i) Azimuthal hadron and target spin angles
\begin{align}
\phi_h(\textrm{present}) = -\phi_p(\textrm{Ref.~\cite{Cosyn:2020kwu}}),
\\
\phi_S(\textrm{present}) = -\phi_S(\textrm{Ref.~\cite{Cosyn:2020kwu}}).
\end{align}

(ii) Angle of target polarization transverse to lepton beam ($\perp$ polarization)
\begin{align}
\psi_N(\textrm{present}) = -\phi_N(\textrm{Ref.~\cite{Cosyn:2020kwu}}).
\end{align}
In the present work we denote the polarization angle around the lepton beam axis by $\psi_N$, to distinguish
it from the collinear-frame angles around the photon axis. We measure the polarization angle in a the same
sense as the collinear frame angles in the Trento convention.

(iii) Transverse basis vectors
\begin{align}
e_{x}(\textrm{present}) = e_{T1}(\textrm{Ref.~\cite{Cosyn:2020kwu}}),
\\
e_{y}(\textrm{present}) = e_{T2}(\textrm{Ref.~\cite{Cosyn:2020kwu}}).
\end{align}

\section{Density matrix from spin operator}
\label{app:density_matrix}
In this appendix we present an alternative derivation of the form of the spin density matrix
of the moving spin-1 system, using the language of quantum-mechanical states and operators.
It starts from the 3-dimensional form of the density matrix in the rest frame and
constructs its 4-dimensional generalization.

The density matrix $\rho$ of a spin--1 system is a 3$\times$3 hermitean matrix 
with unit trace, $\text{Tr}[\rho]=1$. In the rest frame of the spin--1 system, it can be
specified in a basis of single-particle states $|\bm{P} = \bm0;\lambda\rangle$,
where the momentum is zero and the spin is quantized along the $z$-axis, with 
spin projection $\lambda = (+1, 0, -1)$. The density matrix can be parametrized in the form
\cite{Leader:2001gr}
\begin{equation} \label{eq:dens_nonrel}
\rho \equiv \rho(\lambda,\lambda') = 
\frac{1}{3}+\frac{1}{2}S^i\mathscr{S}^i + 
\frac{1}{2}
 T^{ij}(\mathscr{S}^i\mathscr{S}^j+\mathscr{S}^j\mathscr{S}^i - \frac{4}{3}\delta^{ij})\,,
\end{equation}
where $\mathscr{S}^{i (j)}$ are the $3\times 3$ matrices describing the spin 
operators
in the spin--1 representation. In the $|\bm{P} = \bm0;\lambda\rangle$ basis quantized along the $z$-axis,
these take the form
\begin{subequations}
\label{eq:spin_matrices}
\begin{align}
\mathscr{S}^x 
= \frac{1}{\sqrt{2}} \left( \begin{array}{rrr} 0 & \phantom{-}1 & \phantom{-}0 
\\ 1 & 0 & 1 \\ 0 & 1 & 0 \end{array} \right), &
\\
\mathscr{S}^y = 
\frac{i}{\sqrt{2}} \left( \begin{array}{rrr} 0 & -1 & \phantom{-}0 
\\ 1 & 0 & -1 \\ 0 & 1 & 0 \end{array} \right), &
\\
\mathscr{S}^z = 
\left( \begin{array}{rrr} 1 & \phantom{-}0 & \phantom{-}0 
\\ 0 & 0 & 0 \\ 0 & 0 & -1 \end{array} \right) , &
\end{align}
\end{subequations}
and $i,j = (x, y, z)$ denote the Cartesian components. The parameters in 
Eq.~(\ref{eq:dens_nonrel})
are a 3-dimensional pseudovector $S^i$ and a traceless symmetric tensor $T^{ij}$. They coincide, respectively,
with the expectation value of the spin operators and their traceless tensor product
\begin{subequations}
\begin{align}
S^i &= \textrm{Tr} [\rho \hat{\mathscr{S}}^i] = 
\langle\hat{\mathscr{S}}^i\rangle,
\label{eq:nonrel_spin_vector}
\\[1ex]
T^{ij} &= \frac{1}{2}
\textrm{Tr} \left[ \rho 
\left(\hat{\mathscr{S}}^i 
\hat{\mathscr{S}}^j+ \hat{\mathscr{S}}^j \hat{\mathscr{S}}^i 
-\frac{4}{3}\delta^{ij}\right) \right]
\nonumber \\
&= \frac{1}{2}
\left(\langle\hat{\mathscr{S}}^i 
\hat{\mathscr{S}}^j+ \hat{\mathscr{S}}^j \hat{\mathscr{S}}^i
\rangle -\frac{4}{3}\delta^{ij}\right)\,.
\label{eq:nonrel_spin_tensor}
\end{align}
\end{subequations}
In this way the spin--1 density matrix is completely specified by a set of 
expectation values that
can be determined experimentally through spin measurements along certain 
directions.\footnote{This
formulation is analogous to the parametrization of the photon polarization 
density matrix 
through Stokes' parameters~\cite{Berestetskii:1982qgu}.}
Under an active three-rotation $R$ of the rest frame spin--1 system, the parameters $S^{i}$ and $T^{ij}$ 
transform as 
\begin{equation} \label{eq:S_T_rot}
 S^{'i}=R^{ij}S^j, \qquad  T^{'ij}= R^{ik} T^{kl}(R^{-1})^{lj}\,.
\end{equation}
In the rest frame, we can make any choice of axes used in Eq.~(\ref{eq:dens_nonrel}).
In what follows for the application to the semi-inclusive scattering process, we make the choice
corresponding to the ($x'y'z$)-system in Fig.~\ref{fig:kin}.  This is the natural choice, having the
$z$-axis aligned with the collinear axis.  A straightforward longitudinal boost links the target
rest frame and any general collinear frame, and the separation in longitudinal/transverse spin
corresponds to $z$, respectively transverse components of the spin vector $\bm S$ and tensor $T$,
see further.

The parametrization Eq.~(\ref{eq:dens_nonrel}) can be extended to a moving spin--1 system
by replacing the zero--momentum single-particle states by states with momentum $\bm P \neq 0$, 
defined by the action of a standard Lorentz boost $L(P^\mu/M)$ acting on the rest frame states
$|\bm{P} = \bm0;\lambda\rangle$
\begin{equation} \label{eq:std_boost}
 | P,\lambda \rangle \equiv U[L(P^\mu/M)] \;|(M,\bm 0),\lambda 
\rangle \,.
\end{equation}
Here the form of $L(P^\mu/M)$ depends on the choice of dynamics and spin, e.g. canonical, helicity
or light-front helicity~\cite{Keister:1991sb}.

These states transform under a general Lorentz transformation $\Lambda$ with a Wigner rotation
(little group transformation), which depends on the transformation $\Lambda$, particle momentum $P$,
and choice of standard boost $L$:
\begin{equation}
U(\Lambda) | P,\lambda \rangle = \sum_{\lambda'} |\Lambda P,\lambda' \rangle  \, R_w(\Lambda,P/M)_{\lambda'\lambda},
\end{equation}
where the Wigner rotation takes the form
\begin{equation}
 R_w(\Lambda,P/M)\equiv L^{-1}[(\Lambda 
P)^\mu/M]\;\Lambda \;
L(P^\mu/M)\,.\label{eq:wignerrot}
\end{equation}

Invariance of the cross section implies the density matrix $\rho(\lambda,\lambda')$ under general
Lorentz transformations also transforms with Wigner rotations, in similar fashion to
Eq.~(\ref{eq:S_T_rot}) but now using the Wigner rotations of Eq.~(\ref{eq:wignerrot}).  Due to the
transformation properties of the spin operators (Eq.~\ref{eq:spin_matrices}), $S^i$ and $T^{ij}$
then also transform with Wigner rotations.  This leads to a change in the density matrix after a
general Lorentz transformation, where due to the Wigner rotation the values of the matrix elements
depend both on the choice of type of spin and the momentum of the particle.  If the density matrix
parameters are determined through intensity measurements in the ensemble rest frame, the parameters
in any frame where the particle is moving can be obtained by applying Eq.~(\ref{eq:S_T_rot}) using
the relevant Wigner rotation.  The covariant density matrix of Eq.~(\ref{density_tensor}) avoids
these complications as the Wigner rotations cancel between $\rho(\lambda,\lambda')$ and the particle
wave functions (polarization four vectors $\epsilon^\mu(\lambda)$ in the spin-1 case). Overall the
covariant density matrix transforms covariantly, independent of ensemble momentum or type of spin
that is considered.

If one boosts from a reference rest frame to a moving frame using the standard boosts,
Eq.~(\ref{eq:wignerrot}) shows $ R_w(\Lambda,P/M)=1$, and the density matrix in the moving frame is
identical to that of the rest frame.  Between different forms of dynamics, however, different
standard boosts do not result in the same moving frame.  The time-like direction of the reference
rest frame will have been boosted to the same four vector $e_P$, but the spatial directions
corresponding to the boosted $xyz$-axes will be different for different choices of spin.  This
difference is encoded in an additional rest frame rotation, usually referred to as the Melosh
rotation.  If we want to compare density matrices in the \emph{same} moving frame between different
spinor types, we have to include this additional rest frame rotation in the transformations from the
reference rest frame before carrying out the standard boosts. This changes the density matrix,
Eq.~(\ref{eq:S_T_rot}), and yields different density matrices for different choices of spin in the
\emph{same} moving frame (but also different polarization wave functions).

In Sec.~\ref{sec:density_lambda}, the complications related to Wigner rotations are avoided by the
choice made in Eqs.~(\ref{wave_functions_explicit}) to write the polarization basis for the spin-1
target using the coordinate independent basis.  This choice fixes the density matrix for the class
of collinear frames (including a target rest frame) to that of Eq.~(\ref{eq:densitymatrix1}).  These
frames are linked to each other by collinear boosts, which only involve trivial Wigner rotations
($=1$) for the density matrix.  This does not prohibit the use of any frame where the target has
transverse momentum.  The calculations of the elements in Eq.~(\ref{eq:densitymatrix1}) only involve
invariants, which can be computed in any frame.  Its physical interpretation, however, is tied to
the class of collinear frames and happens in coordinate independent fashion through the basis
construction.

Using the moving states of Eq.~(\ref{eq:std_boost}), one can introduce a spin 4-vector and a
symmetric traceless spin 4-tensor, which generalize the 3-dimensional quantities in the rest frame,
Eqs.~(\ref{eq:nonrel_spin_vector}) and (\ref{eq:nonrel_spin_tensor}).  They are defined as the
expectation values
\begin{subequations}
\begin{align}
 S^\mu &= \langle\hat{W}^\mu\rangle\,,\\
  T^{\mu\nu} 
&=
\frac{1}{3}
\langle\hat{W}^\mu 
\hat{W}^\nu + \hat{W}^\nu \hat{W}^\mu + 
\frac{4}{3}\left(g^{\mu\nu}-\frac{\hat{P}^\mu \hat{P}^\nu}{M^2} 
\right)\rangle\,,
\end{align}
\end{subequations}
where 
$\hat{W}^\mu$ is the Pauli-Lubanski operator
\begin{equation}
\hat{W}^\mu=\frac{1}{2M}\epsilon^{\mu\nu\rho\sigma}\hat{M}_{\nu\rho}\hat{P}
_\sigma\,,
\end{equation}
with $\hat{P}^\mu$ and $\hat{M}^{\nu\rho}$ the generators of space-time translations and proper
Lorentz transformations.  It can be shown~\cite{Leader:2001gr} that the $S^\mu$ and $T^{\mu\nu}$
defined as such correspond to the parameters used in the covariant density matrix introduced earlier
in Eq.~(\ref{density_tensor}).

\section{Structure functions as helicity amplitudes}
\label{sec:helicity_amps}
In this appendix we present an alternative derivation of the independent structures in the
spin-1 semi-inclusive cross section using projections of the hadronic tensor on target and virtual photon
helicity states. The projections can be viewed as physical cross sections of the
$\gamma^* + A \rightarrow h + X$ process with definite helicities of the initial particles
(with diagonal and interference terms). The 41 structure functions introduced in
Subsec.~\ref{subsec:structure_functions} can be expressed as linear combinations of the helicity projections.

The target and virtual photon helicity states are defined in a collinear
frame using the $x'y'z$ coordinate system. 
The spin wave functions can be expressed covariantly in terms of the
longitudinal basis vectors of Set II, Eq.~(\ref{basis_longit_set2}), and the transverse
basis vectors aligned with the final hadron momentum, Eq.~(\ref{basis_transv_hadron}).
For the target helicity states
the wave functions are given by Eq.~(\ref{wave_functions_explicit}) and are the same as those
in the spin density matrix Eq.~(\ref{eq:densitymatrix1}).
For the virtual photon, because the 3-momentum is in the negative $z$ direction, the
wave functions are constructed with the $y'$- and $z$-axis flipped \cite{Leader:2001gr},
which leads to
\begin{subequations}
\label{helicity_virtual_photon}
\begin{align}
&\epsilon^\mu (\kappa = 0) = e_{L}^\mu,
\\
&\epsilon^\mu (\kappa = \pm 1) = \mp 
\frac{1}{\sqrt{2}} (e_{x'}^\mu \mp i e_{y'}^\mu) .
\end{align}
\end{subequations}
The helicity projections of the hadronic tensor are then defined as
\begin{equation}
 F^{\lambda'\lambda}_{\kappa' \kappa} \equiv 
\epsilon^{\mu*}(\kappa') W_{\mu\nu}(\lambda',\lambda) \epsilon^\nu (\kappa) \,,
\end{equation}
where $W_{\mu\nu}(\lambda',\lambda)$ is given in Eq.~(\ref{eq:hadronictensor})
and corresponds to the target helicity states Eq.~(\ref{wave_functions_explicit}).
The helicity projections $F^{\lambda'\lambda}_{ij}$ obey the following conditions
due to hermiticity and parity invariance
\begin{subequations}
\begin{align}
&F^{\lambda'\lambda}_{\kappa' \kappa}
=\left( F^{\lambda\lambda'}_{\kappa \kappa'}\right)^*,
\\
&
F^{{-\lambda'}{-\lambda}}_{{-\kappa'}{-\kappa}}
=(-1)^{\lambda-\lambda'+\kappa-\kappa'}F^{\lambda'\lambda}_{\kappa' \kappa}.
\end{align}
\end{subequations}
The conditions reduce the number of independent helicity projections.
Imposing the hermiticity condition we have $3^4=81$ independent real amplitudes;
imposing also the parity invariance condition we are left with 41 (there is
no constraint on the $F^{00}_{00}$ amplitude). Thus the number of independent
real helicity projections is equal to the total number of structure functions
determined by the enumeration of independent structures in Sec.~\ref{subsec:structure_functions}.

The structure functions of Sec.~\ref{subsec:structure_functions} can be expressed as linear
combinations of the helicity projections of the hadronic tensor,
by contracting the independent tensor structures of Eqs.~(\ref{eq:41sf_first})--(\ref{eq:41sf_last})
with the virtual photon helicity wave functions Eq.~(\ref{helicity_virtual_photon}).
We obtain for the unpolarized structure functions
\begin{subequations}
\begin{align}
F_{UU,L} &=\frac{2}{3}
\left(2F^{++}_{00}+F^{00}_{00}\right) ,
\\
F_{UU,T} &=\frac{2}{3}
\left( F^{++}_{++}+F^{00}_{++}+F^{--}_{++}\right),
\\
F_{UU}^{\cos\phi_h} &=-\frac{2\sqrt{2}}{3} 
\operatorname{Re} \left( F^{++}_{+0}+ F^{00}_{+0}+ F^{--}_{+0}\right),
\\
F_{UU}^{\cos2\phi_h} &=-\frac{2}{3} 
\operatorname{Re} \left(2 F^{++}_{+-}+ F^{00}_{+-}\right)\,,
\\
F_{LU}^{\sin\phi_h} &= -\frac{2\sqrt{2}}{3} 
\operatorname{Im} \left( F^{++}_{+0} + F^{00}_{+0} + F^{--}_{+0}\right) ;
\end{align}
\end{subequations}
for the structure functions with target longitudinal vector polarization
\begin{subequations}
\begin{align}
F_{US_L}^{\sin\phi_h} &=-\sqrt{2}\operatorname{Im} \left( F^{++}_{+0} - F^{--}_{+0}\right),
\\
F_{US_L}^{\sin2\phi_h} &=-2\operatorname{Im} F^{++}_{+-},
\\
F_{LS_L} &= F^{++}_{++}-F^{--}_{++},
\\
F_{LS_L}^{\cos\phi_h} &=-\sqrt{2}\operatorname{Re} \left( F^{++}_{+0} - F^{--}_{+0}\right) ;
\end{align}
\end{subequations}
for the structure functions with target transverse vector polarization
\begin{subequations}
\begin{align}
F_{US_T,L}^{\sin(\phi_h-\phi_S)} &= -2\sqrt{2}\operatorname{Im} F^{+0}_{00},
\\
F_{US_T,T}^{\sin(\phi_h-\phi_S)} &= -\sqrt{2}
\operatorname{Re} \left( F^{+0}_{++} - F^{-0}_{++}\right ),
\\
F_{US_T}^{\sin(\phi_h+\phi_S)} &= -\sqrt{2}\operatorname{Im} F^{+0}_{+-},
\\
F_{US_T}^{\sin(3\phi_h-\phi_S)} &= -\sqrt{2}\operatorname{Im} F^{-0}_{+-},
\\
F_{US_T}^{\sin\phi_S} &= -\operatorname{Im} \left( F^{+0}_{+0} +  F^{0-}_{+0} \right),
\\
F_{US_T}^{\sin(2\phi_h-\phi_S)} &= -\operatorname{Im} \left( F^{0+}_{+0} -  F^{-0}_{+0}\right),
\\
F_{LS_T}^{\cos(\phi_h+\phi_S)} &= \sqrt{2}\operatorname{Re}
\left( F^{+0}_{++} + F^{-0}_{++} \right),
\\
F_{LS_T}^{\cos\phi_S} &= -\operatorname{Re} \left( F^{+0}_{+0} + F^{0-}_{+0}\right),
\\
F_{LS_T}^{\cos(2\phi_h-\phi_S)} &= -\operatorname{Re} \left( F^{0+}_{+0} + F^{-0}_{+0}\right);
\end{align}
\end{subequations}
for the structure functions with purely longitudinal tensor target polarization
\begin{subequations}
\begin{align}
 F_{U T_{LL},L} &= 2
\left( F^{++}_{00}-F^{00}_{00}\right),
\\
F_{U T_{LL},T} &= 
\left( F^{++}_{++}-2F^{00}_{++}+F^{--}_{++}\right),
\\
F_{U T_{LL}}^{\cos\phi_h} &= -\sqrt{2}\operatorname{Re}
\left( F^{++}_{+0}-2  F^{00}_{+0}+  
F^{--}_{+0}\right),
\\
F_{U T_{LL}}^{\cos2\phi_h} &= -2 \operatorname{Re}
\left( F^{++}_{+-} - F^{00}_{+-}\right),
\\
F_{L T_{LL}}^{\sin\phi_h} &= -\sqrt{2} \operatorname{Im}
\left( F^{++}_{+0} - 2F^{00}_{+0} + F^{--}_{+0}\right);
\end{align}
\end{subequations}
for the structure functions with longitudinal-transverse tensor polarization
\begin{subequations}
\begin{align}
F_{U T_{LT},L}^{\cos(\phi_h-\phi_{T_L})} &= 4\sqrt{2} \operatorname{Re} F^{+0}_{00},
\\
F_{U T_{LT},T}^{\cos(\phi_h-\phi_{T_L})} &= 2\sqrt{2}\operatorname{Re} \left( F^{+0}_{++} - F^{-0}_{++}\right),
\\
F_{U T_{LT}}^{\cos(\phi_h+\phi_{T_L})} &= -2\sqrt{2} \operatorname{Re} F^{+0}_{+-}, 
\\
F_{U T_{LT}}^{\cos(3\phi_h-\phi_{T_L})} &= 2\sqrt{2} \operatorname{Re} F^{-0}_{+-},
\\
F_{U T_{LT}}^{\cos\phi_{T_L}} &= -2\operatorname{Re} \left( F^{+0}_{+0} - F^{0-}_{+0}\right),
\\
F_{U T_{LT}}^{\cos(2\phi_h-\phi_{T_L})} &= -2\operatorname{Re}
\left( F^{0+}_{+0} - F^{-0}_{+0}\right),
\\
F_{L T_{LT}}^{\sin(\phi_h+\phi_{T_L})} &= -2\sqrt{2}\operatorname{Im}
\left( F^{+0}_{++} + F^{-0}_{++} \right),
\\
F_{L T_{LT}}^{\sin\phi_{T_L}} &= -2 \operatorname{Im} \left( F^{+0}_{+0} - F^{0-}_{+0}\right),
\\
F_{L T_{LT}}^{\sin(2\phi_h-\phi_{T_L})} &= -2\operatorname{Im}
\left( F^{0+}_{+0} - F^{-0}_{+0}\right);
\end{align}
\end{subequations}
and for the structure functions with purely transverse tensor target 
polarization
\begin{subequations}
\begin{align}
F_{U T_{TT},L}^{\cos(2\phi_h-2\phi_{T_T})} &= 2\operatorname{Re} F^{+-}_{00},
\\
F_{U T_{TT},T}^{\cos(2\phi_h-2\phi_{T_T})} &= 2\operatorname{Re} F^{+-}_{++},
\\
F_{U T_{TT}}^{\cos2\phi_{T_T}} &= -\operatorname{Re} F^{+-}_{+-},
\\
F_{U T_{TT}}^{\cos(4\phi_h-2\phi_{T_T})} &= -\operatorname{Re} F^{-+}_{+-},
\\
F_{U T_{TT}}^{\cos(\phi_h-2\phi_{T_T})} &= -\sqrt{2}\operatorname{Re} F^{+-}_{+0},
\\
F_{U T_{TT}}^{\cos(3\phi_h-2\phi_{T_T})} &= -\sqrt{2}\operatorname{Re} F^{-+}_{+0},
\\
F_{L T_{TT}}^{\sin(2\phi_h-2\phi_{T_T})} &= -2\operatorname{Im} F^{+-}_{++},
\\
F_{L T_{TT}}^{\sin(\phi_h-2\phi_{T_T})} &= \operatorname{Im} F^{+-}_{+0},
\\
F_{L T_{TT}}^{\sin(3\phi_h-2\phi_{T_T})} &= -\operatorname{Im} F^{-+}_{+0}.
\end{align}
\end{subequations}
One observes that in each structure function the integer factor before $\phi_h$ is equal to
the difference of the total helicities of the $\gamma^\ast A$ system on the left $(\kappa-\lambda)$
and right side $(\kappa'-\lambda')$ of the projection (the ``initial'' and ``final'' state, if it were
the amplitude of a scattering process),
\begin{align}
m \phi_h, \hspace{2em} |m| = |(\kappa - \lambda) - (\kappa' - \lambda')|.
\end{align}
This expresses the conservation of the angular momentum projection on the $z$-axis.

\section{Extension to higher-spin targets}
\label{app:higher_spin}
In this appendix we outline the extension of the structural decomposition of the
semi-inclusive cross section to targets of spin $>1$.  While there are no
current plans for polarized measurements on targets of higher spin, there are
many atomic nuclei and hadrons with spin $>1$.  From a theoretical viewpoint,
the multipole structure of hadronic matrix elements of QCD operators is also of
general interest~\cite{Cotogno:2019vjb,Cosyn:2025gmp}.  The structure of the
polarization parameters, structure functions and azimuthal dependences follows a
pattern that can be iterated.  A density matrix representing a spin-$j$ ensemble
can be decomposed into $l=0,\ldots,2j$ multipoles of the rotation group.
Compared to the preceding spin $j-1/2$ case, a spin-$j$ target introduces a new
$l=2j$ multipole in the target density matrix with $2(2j)+1$ polarization
parameters ($m_l=-l,\ldots,l$). These parameters can be collected in a symmetric
and traceless rank-$2j$ regular tensor for whole-valued $j$ and pseudotensor for
half-valued $j$.  For the covariant density matrix, these multipole tensors are
Lorentz tensors denoted as $P^{\mu_1\ldots \mu_{2j}}$ here; see the discussion
for spin 1 in Sec.~\ref{subsec:covariant_density_matrix}.  The tensors are
orthogonal to the target four momentum in all their indices.

The polarization parameters that appear in the cross section decomposition can
be written as invariant projections of this tensor on the $L,T$ subspaces
relative to the hadronic scattering plane and collinear axis, see
Eq.~(\ref{eq:poltensor_LT}) for the spin-1 case.  New terms in the decomposition
of the hadronic tensor emerge by combining these invariant polarization
parameters with the independent tensors $\mathcal{A}_i^{\mu\nu}$ and a scalar
structure function.  Only terms that transform as regular tensors are formed,
see Sec.~\ref{subsec:structure_functions}.  The kinematical dependence of the
new terms on $\epsilon$ is determined by the $\mathcal{A}_i^{\mu\nu}$ that
appear in each new term, and their contraction with the leptonic tensor, see
Eqs.~(\ref{eq:L_contraction}).  The following general pattern emerges, which
already became apparent for the spin-1 case:

\begin{enumerate}[(i)]
\item The azimuthal dependence alternates between $\sin$ and $\cos$ and reflects
the parity of the terms.  This follows the pattern

\begin{enumerate}
\item $\cos$-dependence for even parity: $l=\text{even}$ multipole combined with
unpolarized lepton or $l=\text{odd}$ multipole combined with polarized lepton.
\item $\sin$-dependence for odd parity: $l=\text{odd}$ multipole combined with
unpolarized lepton or $l=\text{even}$ multipole combined with polarized lepton.
\end{enumerate}

\item The polarization parameter
\begin{equation}
P_{L\ldots L} = e^{\mu_1}_{L\ast} \ldots e^{\mu_{2j}}_{L\ast} P_{\mu_1\ldots \mu_{2j}},
\end{equation}
which is the projection on the $L$ subspace in all indices, generates five
(whole-valued $j$) or four (half-valued $j$) new structure functions. This is
the $m_l=0$ part of the new multipole in the density matrix.  The azimuthal
dependence of the structure functions is identical to those of the unpolarized
target (five structure functions) or $S_L$ (four) case and only involves multiples of
$\phi_h$ without any offset angles related to polarization.

\item Each projection $P_{L\ldots LT}, P_{L\ldots LTT},\ldots$ on a mixed $L,T$
subspace results in nine new structure functions.  These are the $0< |m_l| <l$
harmonics, where $|m_l|$ correspond to the number of indices that are projected
on the $T$ subspace. The azimuthal and $\epsilon$ dependence follows that from
lower spin multipoles with the same $m_l$ value (same $n\phi_h$ dependence,
offset angle multiplied with $|m_l|$), albeit with a different offset angle
characterizing the higher $l=2j$ multipole and a change between
$\cos/\sin$-dependence if the parity has changed. Compare the $S_T$ and $T_{LT}$
parts in Eqs.~(\ref{cross_section_vector}) and (\ref{cross_section_tensor}) for
an illustration of this pattern in the spin-1 case.

\item $P_{T\ldots T}$, the projection of the polarization tensor on the $T$
subspace in all indices generates nine new structure functions.  This is the
$m_l=\pm l$ multipole. The structure functions carry azimuthal dependencies that
has one more unit of $\phi_h$ compared to the $m_l=l-1=2j-1$ structure
functions, including a new $(2j+2)\phi_h$ modulation, and has a new
$l\phi_{T\ldots T}$ offset angle.  The new terms have a similar dependence on
$\epsilon$ as the preceding $m_l=1,\ldots,l-1$ multipoles.

\item In total the SIDIS cross section for a spin-$j$ target has
\begin{align*}
&5+ (2j \times 9) \qquad (\text{whole-valued}~j),
\nonumber\\
&4+ (2j \times 9) \qquad (\text{half-valued}~j).
\end{align*}
new structure functions compared to the previous spin case. This leads to the
following series for the lowest spin cases
\begin{align*}
5,18,41,72,113,\ldots
\end{align*}
These values can be compared to the counting from helicity projections of the
hadronic tensor, which follows (see Appendix~\ref{sec:helicity_amps})
\begin{align}
\left\lceil \frac{3^2}{2} \times (2j+1)^2 \right\rceil ,
\end{align}
where we used the ceiling function to round to the next integer for whole-valued
$j$.  This is related to parity not providing a constraint on the $F^{00}_{00}$
helicity projection in those cases. One can check that this equation generates
the same series.

\end{enumerate}

\acknowledgments
This study greatly benefited from discussions with A.~Bacchetta, S.~Kumano, and F.~Vera.

This material is based upon work supported by the U.S.~Department of Energy, Office of Science,
Office of Nuclear Physics under contract DE-AC05-06OR23177, and by the U.S.~National Science Foundation
under awards PHY-2111442 and PHY-2239274.

\bibliography{spin1}

\end{document}